%% file: apstemplate.tex
\documentclass[aps,prd,reprint,preprintnumbers,showpacs,floatfix,nofootinbib,superscript address]{revtex4-2}

\usepackage[utf8]{inputenc}
\usepackage{parskip}
\usepackage{amssymb}
\usepackage{stix}
\usepackage{hhline}
\usepackage{amsmath}
\usepackage{mathtools}
\usepackage[dvipsnames]{xcolor}
\usepackage{xspace}
\usepackage{multirow,tabularx}
\usepackage{siunitx}
\usepackage{multirow}
\usepackage{graphicx}
\usepackage{xstring}
\usepackage{etoolbox}
\usepackage{notoccite}
\usepackage{natbib}
\usepackage{mathrsfs}
\usepackage{lineno}
\usepackage{tensor}
\usepackage[thinlines,thicklines]{easybmat}
\usepackage{accents}
\usepackage{tikz}
\allowdisplaybreaks

\input{macros.tex} 

\usepackage{hyperref}
\hypersetup{%
     colorlinks = true,%
     linkcolor = Blue,%
     citecolor = Blue,%
     filecolor = Blue,%
     urlcolor = Blue%
     }%
\usepackage[capitalize]{cleveref} 

\begin{document}

\title{Geometric multipliers and partial teleparallelism in Poincar\'e gauge theory}

\author{W.E.V. Barker}
\email{wb263@cam.ac.uk}
\affiliation{Astrophysics Group, Cavendish Laboratory, JJ Thomson Avenue, Cambridge CB3 0HE, UK}
\affiliation{Kavli Institute for Cosmology, Madingley Road, Cambridge CB3 0HA, UK}


\begin{abstract}
  The dynamics of the torsion-powered teleparallel theory are only viable because thirty-six multiplier fields disable all components of the Riemann--Cartan curvature. 
  We generalise this suggestive approach by considering Poincar\'e gauge theory in which sixty such `geometric multipliers' can be invoked to disable any given irreducible part of the curvature, or indeed the torsion.
  Torsion theories motivated by a weak-field analysis frequently suffer from unwanted dynamics in the strong-field regime, such as the activation of ghosts.
  By considering the propagation of massive, parity-even vector torsion, we explore how geometric multipliers may be able to limit strong-field departures from the weak-field Hamiltonian constraint structure, and consider their tree-level phenomena. 
\end{abstract}

\pacs{04.50.Kd, 04.60.-m, 04.20.Fy}

\maketitle

\section{Introduction}\label{section_26a}

The Poincar\'e gauge theory of gravity (PGT), as pioneered by Kibble~\cite{1961JMP.....2..212K}, Sciama~\cite{RevModPhys.36.463}, Utiyama~\cite{PhysRev.101.1597}, and many others~\cite{Hehl:1976kj,blagojevic2002gravitation,Obukhov:2006gea,1998RSPTA.356..487L}, naturally extends general relativity (GR) as a diffeomorphism gauge theory, by additionally gauging the Lorentz group $\soonethree$. This innovation allows for spacetime torsion, which may be dynamical (and thus engender new phenomena~\cite{2005NewAR..49...59P,2011PhRvD..83b4001B,chapter3}), and possibly allows for a breaking of the equivalence principle through the details of the gravitational coupling to the various $\sltwoc$ representations of matter~\cite{2014IJMPD..2342004P,2010RPPh...73e6901N,2016IJMPS..4060010N}.

The basic units of the `\emph{particle physics}' formulation of PGT~\cite{blagojevic2002gravitation,chapter2,chapter3,chapter4,mythesis} are the translational and rotational gauge fields $\tensor{b}{^i_\mu}$ and $\tensor{A}{^{ij}_\mu}=\tensor{A}{^{[ij]}_\mu}$, where Greek indices are holonomic, referring to coordinates on a flat and torsion-free reference background $\check{\mathcal{M}}$, and Roman indices are Lorentzian. In the alternative `\emph{geometric}' formulation, these gauge fields are reimagined as the tetrad $\tensor{e}{^i_\mu}$ and spin-connection $\tensor{\omega}{^{ij}_{\mu}}\equiv\tensor{\omega}{^{[ij]}_{\mu}}$, while the spacetime enjoys an intrinsic curvature and torsion~\cite{2019Univ....5..173B,blagojevic2002gravitation}. In the former interpretation -- which we use here only out of convenience -- these geometric quantities are interpreted as the field strength tensors 
\begin{subequations}
  \begin{align}
    \tensor{\mathcal{R}}{^{ij}_{kl}}&\equiv 2\tensor{h}{_k^\mu}\tensor{h}{_l^\nu}\big(\tensor{\partial}{_{[\mu|}}\tensor{A}{^{ij}_{|\nu]}}+\tensor{A}{^i_{m[\mu|}}\tensor{A}{^{mj}_{|\nu]}}\big),\label{riemanndef}
    \\
    \tensor{\mathcal{T}}{^i_{kl}}&\equiv 2\tensor{h}{_k^\mu}\tensor{h}{_l^\nu}\big(\tensor{\partial}{_{[\mu|}}\tensor{b}{^i_{|\nu]}}+\tensor{A}{^i_{m[\mu|}}\tensor{b}{^m_{|\nu]}}\big)\label{torsiondef}.
\end{align}
\end{subequations}
In terms of these tensors, it is common to construct, for Planck mass $\planck$, the \emph{quadratic} version of PGT
\begin{equation}
  \begin{aligned}
    \tensor{L}{_{\text{G}}}&=
    -\frac{1}{2}\alp{0}\planck^2\mathcal{R}
    +\sum_{I=1}^{6}\tensor{\hat{\alpha}}{_I}\tensor{\mathcal{  R}}{^{ij}_{kl}}
				    \projlore[_{ij}^{kl}_{nm}^{pq}]{I}\tensor{\mathcal{  R}}{^{nm}_{pq}}\\
				    &\ \ +\planck^2\sum_{M=1}^{3}\tensor{\hat{\beta}}{_M}\tensor{\mathcal{  T}}{^{i}_{jk}} 
				    \projlore[_{i}^{jk}_{l}^{nm}]{M}\tensor{\mathcal{  T}}{^{l}_{nm}},
  \label{pgtqp}
  \end{aligned}
\end{equation}
i.e. extending the Einstein--Cartan--Kibble--Sciama (ECKS) theory~\cite{2006gr.qc.....6062T} by the collection of possible Maxwell-like terms via $\{\alp{I}\}$, $\{\bet{M}\}$, which could be used to introduce dynamical torsion. The $\projlore[_{ij}^{kl}_{nm}^{pq}]{I}$ and $\projlore[_{i}^{jk}_{l}^{nm}]{M}$ project out the irreducible Lorentz group representations which are contained within the field strength tensors. As some more recent authors have noted~\cite{2018PhRvD..98b4014B,2011PhRvD..83b4001B,Diakonov:2011fs,2011IJMPD..20.2125H,2011CQGra..28u5017B}, invariants may be written down beyond the nine considered above, if an extension is made to mixed-parity Lagrangia. We make no physical case for excluding such terms, but restrict our discussion to the parity-even PGT theory\footnote{Note that in~\cite{chapter2,chapter3,chapter4,mythesis} we are refering to the parity-even, quadratic action in~\eqref{pgtqp} as PGT\textsuperscript{q,+} rather than PGT.}.

If $\alp{0}=1$, we interpret~\eqref{pgtqp} as a `\emph{modified gravity}' theory which deviates quadratically from ECSK, and so phenomenologically from GR. For some time the real concern has been promulgated~\cite{1999IJMPD...8..459Y,2002IJMPD..11..747Y,2018PhRvD..98b4014B,BeltranJimenez:2019hrm} that only two special cases of this configuration may be viable: those with additional even/odd-parity $0^+$ or $0^-$ scalar torsion modes (i.e. degrees of freedom, d.o.f)~\cite{1999IJMPD...8..459Y}. In general, the parameters of~\eqref{pgtqp} also allow for $1^+$, $1^-$, $2^+$ and $2^-$ modes, but apparently no single one of these can be invoked within the linear regime (i.e. near a flat, torsion-free spacetime), without activating others in the full nonlinear theory~\cite{2002IJMPD..11..747Y,1998AcPPB..29..961C}. These uninvited modes are thought of as \emph{strongly coupled}, i.e. becoming non-dynamical on the Minkowski background. Whether strong coupling is intrinsically problematic can only be determined in detail for a given theory (see~\cite{crunchy} and references therein). In the case of the PGT, however, the unitarity conditions (of the linearised theory) on the $\{\alp{I}\}$, $\{\bet{M}\}$, inevitably cause the nonlinearly activated modes to contribute negative energies in the Hamiltonian. To whatever extent this is true, strong coupling will be fatal to the more general PGTs. We confirmed in~\cite{chapter4} that the problem extends also into linearly viable cases for which $\alp{0}=0$.

\vspace{10pt}

In this paper, therefore, we seek an extension of the PGT which tends to ameliorate the nonlinear proliferation of propagating d.o.f.
We will first require this extension to be \emph{minimal}. Many attractive options present, for example, when one considers alternatives to the Poincar{\'e} gauge group $\poincare$. 
However, since these alternatives are chiefly realised within existing frameworks such as \emph{Weyl} gauge theory WGT~\cite{1973PThPh..50.2080U,1975PThPh..53..565U}, its \emph{extended} alternative (eWGT)~\cite{lasenby-hobson-2016} and the \emph{metric-affine} generalisation (MAGT)~\cite{1988PhLB..200..489N}, they do not lie comfortably within our scope. 
Nor shall we augment the PGT with new (manifestly) dynamical fields, such as the scalar added by Horndeski to Einstein's theory~\cite{1974IJTP...10..363H}. 
In fact a particularly conservative approach is suggesed already within PGT, in the form of \emph{teleparallel gravity}~\cite{2000CQGra..17.3785B}.
The teleparallel form of GR has total Lagrangian\footnote{Note that our conventions for multipliers, which we take to be tensors, will differ from those used in~\cite{blagojevic2002gravitation}, where they are treated as densities.}
\begin{equation}
\begin{aligned}
  \lagrangian{T}&=\frac{1}{2}{m_{\text{p}}}^2\mathbb{T}+\tensor{\lambda}{_{ij}^{kl}}\tensor{\mathcal{  R}}{^{ij}_{kl}}+L_{\text{M}}, \\
  \mathbb{T}&\equiv\frac{1}{4}\mathcal{  T}_{ijk}\mathcal{  T}^{ijk}+\frac{1}{2}\mathcal{  T}_{ijk}\mathcal{  T}^{jik}-\mathcal{  T}_i\mathcal{  T}^i.
  \label{fullteleparallel}
\end{aligned}
\end{equation}
The dynamical part $\mathbb{T}$ of~\eqref{fullteleparallel} is purely quadratic in torsion. There is however added to this Lagrangian a kinematic term, which suppresses the whole Riemann--Cartan curvature by means of 36 multiplier fields $\tensor{\lambda}{_{ij}^{kl}}$. 
In the geometric interpretation, the multipliers constrain the rich Riemann--Cartan geometry $U_4$ to that of Weitzenb\"ock $T_4$, eliminating unwanted modes in the process.
They do not however appear as propagating d.o.f in the final counting: those that persist in the equations of motion (e.o.m) do so as \emph{determined} quantities, so that those same equations do not add physical content while the rest only propagate the $2^+$ graviton.
In the grand picture, the theory~\eqref{fullteleparallel} is rightly considered a PGT, since the multipliers play a \emph{restrictive} role. 
In this paper we therefore focus on \emph{general} geometry-constraining multipliers in the PGT context, with an intended application to strong coupling.

The remainder of this paper is set out as follows. In~\crefrange{formalism}{graveqsn} we set out the general theory of `geometric multipliers' in the Lagrangian formulation. We consider the new, general Hamiltonian structure in~\crefrange{conpri}{consec}, indicating the mechanism by which the multipliers may help to soften the dynamical transition from linear to nonlinear gravity. In~\cref{lorspe} we perform the canonical analysis of the $1^+$ torsional mode with a simple choice of multiplier. Conclusions follow in~\cref{chapter5conclusions}. Most of our conventions and notation, especially for the Hamiltonian, ADM or $3+1$ structure of the PGT, are in common with~\cite{chapter4,mythesis}, and with the companion paper in~\cite{crunchy}; these conventions are in turn derived from~\cite{blagojevic2002gravitation,1999IJMPD...8..459Y,2002IJMPD..11..747Y}. We use the `West Coast' signature $(+,-,-,-)$.

\section{The Lagrangian picture}\label{lagpic}

The subject of our investigation is the covariant restriction of the Riemann--Cartan geometry through the introduction of \emph{geometric multipliers}. An additional 60 gravitational d.o.f are added to the PGT via the multiplier fields $\tensor{\lambda}{^i_{jk}}$ and $\tensor{\lambda}{^{ij}_{kl}}$, which share the symmetries and dimensions of the Riemann--Cartan and torsion tensors. The new gravitational Lagrangian which replaces~\eqref{pgtqp} is written as
  \begin{align}
    \tensor{L}{_{\text{G}}}&=
    -\frac{1}{2}\alp{0}\planck^2\mathcal{R}
    \nonumber\\
    &\ \ +\sum_{I=1}^{6}\Big(\tensor{\hat{\alpha}}{_I}\tensor{\mathcal{  R}}{^{ij}_{kl}}
				    +\tensor{\bar{\alpha}}{_I}\tensor{\lambda}{^{ij}_{kl}}\Big)
				    \projlore[_{ij}^{kl}_{nm}^{pq}]{I}\tensor{\mathcal{  R}}{^{nm}_{pq}}\nonumber\\
				    &\ \ +\planck^2\sum_{M=1}^{3}\Big(\tensor{\hat{\beta}}{_M}\tensor{\mathcal{  T}}{^{i}_{jk}}
				    +\tensor{\bar{\beta}}{_M}\tensor{\lambda}{^{i}_{jk}}\Big)
				    \projlore[_{i}^{jk}_{l}^{nm}]{M}\tensor{\mathcal{  T}}{^{l}_{nm}},
  \label{neocon}
  \end{align}
where any nonvanishing $\{\calp{I}\}$ and $\{\cbet{M}\}$ switch off the various irreducible representations of $\soonethree$ which are contained within the field strengths: the bar indicates coefficients associated with the Lagrange multiplier tensor fields. The $2^3\times 2^6$ configurations of these boolean `switches' allow the greatest possible control over theory beyond the nine `dials' which define the original PGT, whilst maintaining general covariance.

\subsection{Simple toy model}\label{SimpleToyModel}

The apparent complexity of the Dirac Hamiltonian constraint algorithm, when applied in its most involved form to higher-rank field theories with various spin sectors, and with variational interpretation of the Poisson brackets, can obfuscate the simplicity of the physics involved. In the good pedagogical introductions to the algorithm, field theory is not actually used in the first instance, rather it is sufficient to analyse simple, one-dimensional `beads on a string' examples.

Accordingly, we give a rough example in this section of how a multiplier field can extend the nonlinear dynamics down onto a pathological phase surface, with a `toy model' (TM) system 
\begin{equation}\label{TM}
	L_{\text{TM}}\equiv\dot{q}_1q_2\left(1+q_2\right)+\dot{q}_2q_1\left(1+q_1\right)+\lambda\left(\dot{q}_1+\dot{q}_2\right).
\end{equation}
In this example, $q_1$ and $q_2$ are real, generalised coordinates, the dot denotes a derivative with respect to time $t$, and there will be a corresponding action $S_{\text{TM}}\equiv\int\mathrm{d}tL_{\text{TM}}$. Note that there is also a real multiplier field, $\lambda$. The Lagrangian looks a little peculiar, because there are fewer derivatives than expected in a physical system. For our purposes, we do not need to worry about this: the collection of interactions in~\cref{TM} is of the kind that could well arise among many others in a gauge gravity theory, once the spin-parity decomposition is performed. For the toy model, we imagine that anomalous dimensionality among the operators can be accounted for by (i) even powers of the Planck mass, which we omit (ii) interpretation of either $q_1$ or $q_2$ as dimensionful connection fields or dimensionless metric (or tetrad) potentials and (iii) recalling that in the field theory (but not in the one-dimensional model) we have spatial gradients which contribute a mass dimension without (modulo important boundary effects) affecting the dynamics on the same footing as time derivatives. We will take the `vacuum spacetime' to simply be $q_1\approx q_2\approx 0$, which is an exact solution to the field equations of~\cref{TM} whether or not the multiplier is included. Near this vacuum, there will be a \emph{linearised} theory where the Lagrangian is expanded up to second order in perturbative fields. Both the linearised and fully nonlinear theories are subject to Hamiltonian analysis, and the results had better be consistent. 

First try omitting the multiplier: the definitions $\pi_i\equiv\partial/\partial\dot{q}_i L_{\text{TM}}\left(\lambda\to 0\right)$ naturally engender two \emph{primary} constraints $\phi_1\equiv\pi_1-q_2\left(1+q_2\right)\approx 0$ and $\phi_2\equiv\pi_2-q_1\left(1+q_1\right)\approx 0$, so the Hamiltonian is $H_{\text{TM}}\equiv\sum_i u_i\phi_i\approx 0$, where canonical variables $u_i$ are Dirac's `missing/uninvertible velocities' $\dot{q}_i$. Now the Poisson bracket is $\left\{\phi_1,\phi_2\right\}\approx 2\left(q_1-q_2\right)$, which has the sickly property that it \emph{vanishes} in the linearised theory. For vanishing Poisson bracket, the consistency conditions $\dot{\phi}_i\equiv\left\{\phi_i,H_{\text{TM}}\right\}=\sum_{i}\left\{\phi_i,\phi_j\right\}u_j\approx0$ are automatically satisfied, and the undetermined $u_i$ embody the pure gauge freedoms of $q_i$ in the trivial (in fact \emph{total derivative}) linearisation $L_{\text{TM}}\left(\lambda\to 0\right)=\mathrm{d}/\mathrm{d}t\left(q_1q_2\right)+\mathcal{O}\left(q^3\right)$. More formally there are $1+1=2$ na\"ive d.o.f $q_i$, but $\phi_i$ are \emph{first class}, so $2-\frac{1}{2}2\times 2=0$ non-gauge d.o.f propagate according to Dirac algorithm `lore'~\cite{Golovnev:2022rui}. For a nonvanishing bracket (i.e. nonlinear theory), the consistency conditions demand $u_i\approx 0$, while $\phi_i$ become \emph{second class}, so $2-\frac{1}{2}1\times 2=1$ d.o.f propagates. Indeed, the nonlinear Euler--Lagrange equations are $\dot{q}_i\approx 0$, so the `lore' interprets the two initial data $q_i(t=0)$ as the Cauchy data for \emph{one} effective oscillator d.o.f, which vanishes in the linear spectrum\footnote{In our toy model of course, there is no harmonic oscillator, so to complete the analogy an example with more derivatives and more `spectator' fields representing the other spin sectors should really be constructed so as to make the usual constrained d.o.f counting interpretation strictly accurate. This setup would be less minimal than the one in~\cref{TM}.}. Let's explicitly check that this nonlinear result is consistent with the nonlinear Euler--Lagrange equations without the multiplier;
\begin{equation}\label{NoMultiplier}
	\dot{q}_2\left(q_1-q_2\right)\approx\dot{q}_1\left(q_2-q_1\right)\approx 0.
\end{equation}
Yes: the $u_i$ vanish on shell and so, correspondingly, do the $\dot{q}_i$ according to~\cref{NoMultiplier}. There may be an exception if $q_1\approx q_2$, but again this corresponds to the bracket \emph{vanishing}, so it is just another instance of the same problem. In general, we are most immediately concerned with brackets which vanish upon departure from preferred spacetimes (such as our vacuum, or the Minkowski solution). The effect may punctuate the bulk of the phase space elsewhere, but it is then a question of whether such spacetimes are really observed in nature. Returning to the linear case, we see that the field equations~\cref{NoMultiplier} contain no first-order terms, so they completely evaporate under linearisation. In the bulk, we needed to fix the constant $q_i$ with our initial data, but in the linearisation the $q_i$ are completely arbitrary. In effect, a very trivial gauge symmetry in the linear theory (arbitrary, time-local transformations of both fields) is nonlinearly broken. This makes the linear physics suspicious and untrustworthy, however appealing it might be from a particle spectrum perspective. The breaking of the gauge symmetry is actually not the key feature we wish to capture: rather it is the loss of propagating d.o.f. It is straightforward to also construct one-dimensional examples akin to~\cref{TM} where the dissapearing d.o.f is due to a proliferation of secondary constraints rather than a change in class of the primaries --- we will not do so here, but both mechanisms can be reaslised in the prolific tangle of gauge gravity interactions seen in the PGT, and both mechanisms have at their heart a dissapearing Poisson bracket.

Now $\lambda$ in~\cref{TM} increases us to $2+1=3$ na\"{i}ve d.o.f: it is easy to check that the first class combination $\phi_-\equiv\phi_1-\phi_2+2(q_1-q_2)\phi_\lambda\approx 0$ and second class $\phi_+\equiv\phi_1+\phi_2\approx 0$ and $\phi_\lambda\equiv\pi_\lambda\approx 0$ with linear/nonlinear nonvanishing bracket $\left\{\phi_+,\phi_\lambda\right\}\approx -2$ indicate $3-\frac{1}{2}\left(2\times 1+1\times 2\right)=1$ d.o.f. Again, let's check that this really works by extending the field equations from~\cref{NoMultiplier} with the presence of the multiplier;
\begin{equation}\label{WithMultiplier}
	\dot{q}_2\left(q_1-q_2\right)\approx\dot{q}_1\left(q_2-q_1\right)\approx \dot{\lambda}/2,\quad \dot{q}_1+\dot{q}_2\approx 0.
\end{equation}
So we see from~\cref{WithMultiplier} how the situation develops. In the nonlinear case, two initial data such as $q_1(t=0)+q_2(t=0)$ and $\lambda(t=0)$ can be propagated, but for the former coordinate combination we can only do this once a single extra pure gauge variable such as $q_1-q_2$ has been specified along the phase trajectory. This is similar to the \emph{nonlinear} scenario without the multiplier. There is an extra gauge variable, but once again it is possible to construct other toy examples of this mechanism where the comparison is exact. Now the really interesting feature of~\cref{WithMultiplier} is that the equations of motion (and their spectral ramifications) \emph{persist} in the linear theory: they now have terms first-order in the generalised coordinates, and it is these key terms (velocities) which dictate the dynamics. In the linearised case, the gauge coordinate is not needed to propagate either of the others, but since it is present in the Lagrangian it must still be specified to fully evolve the system -- thus the dynamics are unchanged. The overall effect is to drag the nonlinear dynamics down into the linear regime. This opens the door to solving a major problem in non-Riemannian gravity, which we explore through the rest of this work.

\subsection{Developing the formalism}\label{formalism}

In order to efficiently and thoroughly discuss the new general theory~\eqref{neocon}, we must create a more formal notation than that previously used in~\cite{chapter4}.
We use (as in~\cref{pgtqp,neocon}) the indices $I$, $J$, $K$ and $L$ to label the $\soonethree$ irreps of $\tensor{\mathcal{  R}}{^{ij}_{kl}}$, ranging from one to six, and we also allocate $M$, $N$, $O$ and $P$ to label those of $\tensor{\mathcal{  T}}{^i_{jk}}$, ranging from one to three. 

The Arnowitt--Deser--Misner (ADM) split then allows us to construct something similar using the rotation group $\sothree$. There is a spacelike slicing, characterised by a unit timelike vector $\tensor{n}{_k}$ which can be extracted from the translational gauge field as follows
\begin{equation}
	\tensor{n}{_k}\equiv\tensor{h}{_k^0}/\sqrt{\tensor{g}{^{00}}},
	\label{foliation_master}
\end{equation}
where the clock-and-ruler metric elements are recovered in PGT by $\tensor{g}{^{\mu\nu}}\equiv\tensor{h}{_i^\mu}\tensor{h}{_j^\nu}\tensor{\eta}{^{ij}}$.
With respect to this vector, any indexed quantity can then be split into perpendicular and parallel parts ${\tensor{\mathcal{  V}}{^i}=\tensor{\mathcal{  V}}{^\perp}\tensor{n}{^i}+\tensor{\mathcal{  V}}{^{\ovl i}}}$. Note some further identities ${\tensor{b}{^{\ovl k}_\alpha}\tensor{h}{_{\ovl{l}}^\alpha}\equiv\tensor*{\delta}{_{\ovl l}^{\ovl k}}}$ and ${\tensor{b}{^{\ovl k}_\alpha}\tensor{h}{_{\ovl{k}}^\beta}\equiv\tensor*{\delta}{_\alpha^\beta}}$. As set out in~\cite{chapter4}, the overbar on an index, and the ($\perp$) symbol, refer to indices perpendicular and parallel to the ADM unit vector $\foli{i}$. For a discussion of the ADM formulation of PGT with these exact same conventions, see~\cite{chapter4,blagojevic2002gravitation}.

Using~\eqref{foliation_master} we now also introduce $A$, $B$, $C$ and $D$ to span the $\othree$ irreps in the \emph{rotational context} -- such as those contained within $\tensor{\hat{\pi}}{_{ij}^{\ovl{k}}}$, $\tensor{\mathcal{  R}}{^{ij}_{\ovl{kl}}}$ and $\tensor{\mathcal{  R}}{^{ij}_{\perp\ovl{l}}}$ -- which are $0^+$, $0^-$, $1^+$, $1^-$, $2^+$ and $2^-$. 
The parallel momentum $\smash{\tensor{\hat{\pi}}{_{ij}^{\overline{k}}}\equiv\tensor{\pi}{_{ij}^\alpha}\tensor{b}{^{k}_\alpha}}$, where $\alpha$, $\beta$, etc. exclude the time index, refer to the rotational momentum $\tensor{\pi}{_{ij}^\mu}$ conjugate to $\tensor{A}{^{ij}_{\mu}}$. 

We will use $E$, $F$, $G$ and $H$ to span these \emph{same} irreps in the \emph{translational context}, i.e. wherever such irreps are present in $\tensor{\hat{\pi}}{_{i}^{\ovl{k}}}$ (the parallel part of the $\tensor{b}{^i_\mu}$ conjugate momentum $\tensor{\pi}{_i^\mu}$), $\tensor{\mathcal{  T}}{^{i}_{\ovl{kl}}}$ and $\tensor{\mathcal{  T}}{^{i}_{\perp\ovl{l}}}$. Care must be taken, since various of the six spin-parity ($J^P$) irreps are missing from various objects in the translational sector (there are no $0^-$ or $2^-$ parts in the $\tensor{\pi}{_i^{\ovl{k}}}$ or $\tensor{\mathcal{T}}{^i_{\perp\ovl{j}}}$, and no $0^+$ or $2^+$ parts in $\tensor{\mathcal{T}}{^i_{\ovl{jk}}}$), and summations over the new indices are assumed to take this into account implicitly.

Using this notation, we next introduce the `human readable' projections as denoted with a h\'a\v{c}ek ($\check{\cdot}$) $\tensor[^A]{\hat{\pi}}{_{\acu{l}}}\equiv\projorthhum[{_{\acu{l}}^{ij}_{\ovl{k}}}]{A}\tensor{\hat{\pi}}{_{ij}^{\ovl{k}}}$, $\tensor[^E]{\hat{\pi}}{_{\acu{l}}}\equiv\projorthhum[{_{\acu{l}}^{i}_{\ovl{k}}}]{E}\tensor{\hat{\pi}}{_{i}^{\ovl{k}}}$, etc.
These obtain the convenient variable-index expressions of the $J^P$ parts, such as $\PiP{A0p}$, $\PiP{A1p}$, $\PiP{A2p}$ etc., (respectively $0^+$, $1^+$ and $2^+$) as used previously in~\cite{chapter4,mythesis,2002IJMPD..11..747Y}, and where a \emph{variable} number of indices\footnote{See Lin's notation in~\cite{Lin1}} is denoted by $\acu{u}$, $\acu{v}$, $\acu{w}$, etc. We provide a full list of variable-index expressions in~\cref{ireppl}, and use both the formal $A$ and $E$ notation and the older variable-index notation interchangeably.
To account for missing irreps, we define placeholder projections within the translational sector
\begin{equation}
  \projorthhum[{_{\acu{v}}^{i}_{\ovl{jk}}}]{ {0^+}}\equiv
\projorthhum[{_{\acu{v}}^{i}_{\ovl{jk}}}]{ {2^+}}\equiv
\projorthhum[{_{\acu{v}}^{i}_{\ovl{k}}}]{ {0^-}}\equiv
  \projorthhum[{_{\acu{v}}^{i}_{\ovl{k}}}]{ {2^-}}\equiv
  0.
  \label{plahol}
\end{equation}

There is a corresponding complete (i.e. not variable-index) set of operators which is denoted with a circumflex ($\hat{\cdot}$). It is convenient to describe relations between both sets of operators using the dimensionless numbers $\{\ctmp[A]{\parallel}\}$, $\{\ctmp[E]{\parallel}\}$, $\{\ctmp[A]{\perp}\}$, $\{\ctmp[E]{\perp}\}$, which are close to unity
  \begin{align}
    \projorth[_{ij}^{\ovl{kp}}_{lm}^{\ovl{nq}}]{A}&\equiv\ctmp[A]{\parallel}\projorthhumu[^{\acu{u}}_{ij}^{\ovl{kp}}]{A}\projorthhum[_{\acu{u}}_{lm}^{\ovl{nq}}]{A},\quad
    \tensor*{\delta}{_{\acu{u}}^{\acu{v}}}\equiv\ctmp[A]{\parallel}\projorthhumu[_{\acu{u}}_{ij}^{\ovl{kp}}]{A}\projorthhum[^{\acu{v}}^{ij}_{\ovl{kp}}]{A},\nonumber\\
\projorth[_{ij}^{\ovl{k}}_{lm}^{\ovl{n}}]{A}&\equiv\ctmp[A]{\perp}\projorthhumu[^{\acu{u}}_{ij}^{\ovl{k}}]{A}\projorthhum[_{\acu{u}}_{lm}^{\ovl{n}}]{A},\quad
    \tensor*{\delta}{_{\acu{u}}^{\acu{v}}}\equiv\ctmp[A]{\perp}\projorthhumu[_{\acu{u}}_{ij}^{\ovl{p}}]{A}\projorthhum[^{\acu{v}}^{ij}_{\ovl{p}}]{A},\nonumber\\
    \projorth[_{i}^{\ovl{kp}}_{l}^{\ovl{nq}}]{E}&\equiv\ctmp[E]{\parallel}\projorthhumu[^{\acu{u}}_{i}^{\ovl{kp}}]{E}\projorthhum[_{\acu{u}}_{l}^{\ovl{nq}}]{E},\quad
    \tensor*{\delta}{_{\acu{u}}^{\acu{v}}}\equiv\ctmp[E]{\parallel}\projorthhumu[_{\acu{u}}_{i}^{\ovl{kp}}]{E}\projorthhum[^{\acu{v}}^{i}_{\ovl{kp}}]{E},\nonumber\\
\projorth[_{i}^{\ovl{k}}_{l}^{\ovl{n}}]{E}&\equiv\ctmp[E]{\perp}\projorthhumu[^{\acu{u}}_{i}^{\ovl{k}}]{E}\projorthhum[_{\acu{u}}_{l}^{\ovl{n}}]{E},\quad
    \tensor*{\delta}{_{\acu{u}}^{\acu{v}}}\equiv\ctmp[E]{\perp}\projorthhumu[_{\acu{u}}_{i}^{\ovl{p}}]{E}\projorthhum[^{\acu{v}}^{i}_{\ovl{p}}]{E}.
\end{align}
These complete operators are more cumbersome in their actual form, but useful for formal calculations. 

Most importantly, we introduce a compact notation for the linear combinations of coupling constants which will arise frequently at all levels of analysis.
Accordingly there are eight matrices, again populated by numbers close to unity
\begin{widetext}
  \begin{align}
    \projorthhum[_{\acu{p}}^{lm}_{\ovl{nq}}]{A}\projlore[_{lm}^{\ovl{nq}}_{ij}^{\ovl{rk}}]{I}&\equiv \projmatrix[AI]{\parallel\parallel}\projorthhum[_{\acu{p}}_{ij}^{\ovl{rk}}]{A},\quad
    \projorthhum[_{\acu{p}}^{lm}_{\ovl{n}}]{A}\projlore[_{lm}^{\perp\ovl{n}}_{ij}^{\ovl{rk}}]{I} \equiv \projmatrix[AI]{\perp\parallel}\projorthhum[_{\acu{p}}_{ij}^{\ovl{rk}}]{A},\quad
    \projorthhum[_{\acu{p}}^{lm}_{\ovl{nq}}]{A}\projlore[_{lm}^{\ovl{nq}}_{ij}^{\perp\ovl{k}}]{I} \equiv \projmatrix[AI]{\parallel\perp}\projorthhum[_{\acu{p}}_{ij}^{\ovl{k}}]{A},\nonumber\\
    \projorthhum[_{\acu{p}}^{lm}_{\ovl{n}}]{A}\projlore[_{lm}^{\perp\ovl{n}}_{ij}^{\perp\ovl{k}}]{I} &\equiv \projmatrix[AI]{\perp\perp}\projorthhum[_{\acu{p}}_{ij}^{\ovl{k}}]{A},\quad
    \projorthhum[_{\acu{p}}^{l}_{\ovl{nq}}]{E}\projlore[_{l}^{\ovl{nq}}_{i}^{\ovl{rk}}]{M} \equiv \projmatrix[EM]{\parallel\parallel}\projorthhum[_{\acu{p}}_{i}^{\ovl{rk}}]{E},\quad
    \projorthhum[_{\acu{p}}^{l}_{\ovl{q}}]{E}\projlore[_{l}^{\perp\ovl{q}}_{i}^{\ovl{rk}}]{M} \equiv \projmatrix[EM]{\perp\parallel}\projorthhum[_{\acu{p}}_{i}^{\ovl{rk}}]{E},\nonumber\\
    \projorthhum[_{\acu{p}}^{l}_{\ovl{nq}}]{E}\projlore[_{l}^{\ovl{nq}}_{i}^{\perp\ovl{k}}]{M} &\equiv \projmatrix[EM]{\parallel\perp}\projorthhum[_{\acu{p}}_{i}^{\perp\ovl{k}}]{E},\quad
    \projorthhum[_{\acu{p}}^{l}_{\ovl{q}}]{E}\projlore[_{l}^{\perp\ovl{q}}_{i}^{\perp\ovl{k}}]{M} \equiv \projmatrix[EM]{\perp\perp}\projorthhum[_{\acu{p}}_{i}^{\perp\ovl{k}}]{E},
    \label{manymany}
  \end{align}
\end{widetext}
which encode the \emph{transfer} of $\othree$ projections \emph{through} the $\soonethree$ projections. With these matrices we obtain various \emph{transfer couplings}, using the obvious notation ${\alpm[A]{\parallel\parallel}\equiv\sum_{I}\projmatrix[AI]{\parallel\parallel}\alp{I}}$, ${\cbetm[E]{\perp\parallel}\equiv\sum_{M}\projmatrix[EM]{\perp\parallel}\cbet{M}}$, etc.
Due to~\eqref{plahol}, the relations~\eqref{manymany} do not fully define these quantities and we again supplement with the \emph{vanishing} placeholder couplings $\cbetm[{0^+}]{\perp\parallel}$, $\cbetm[{2^+}]{\perp\parallel}$, $\cbetm[{0^-}]{\parallel\perp}$, $\cbetm[{2^-}]{\parallel\perp}$, $\betm[{0^+}]{\perp\parallel}$, $\betm[{2^+}]{\perp\parallel}$, $\betm[{0^-}]{\parallel\perp}$ and $\betm[{2^-}]{\parallel\perp}$.
Explicit formulae for all transfer couplings are provided in~\cref{mulcou}.
We shall show in~\cref{hampiq} that the canoncial structure of PGT and the geometric multiplier extension in~\eqref{neocon} can be fully understood through the transfer couplings and their relations~\cref{doubleplusgood,plusgood}.

Finally, we will add two more items of formalism by defining the functions
  \begin{align}
     \mu(x) \equiv \left\{\begin{array}{lr}
	 x^{-1}, & \text{for } x\neq 0\\
	 0, & \text{for } x=0,
     \end{array}\right.\quad
     \nu(x) \equiv 1-|\text{sgn}(\mu(x))|.
\end{align}
These functions allow for a general discussion of constrained quantities in the Hamiltonian picture, and in particular the function $\mu(x)$ is not new, being defined already by Blagojevi\'c and Nikoli\'c in~\cite{1983PhRvD..28.2455B}, as part of the crucial \emph{if-constraint formalism}. An if-constraint is a Hamiltonian constraint which appears only because the couplings in~\eqref{pgtqp} obey certain critical relations; we will be using this formalism in~\cref{hampiq} when we address the Hamiltonian picture.

\subsection{The gravitational field equations}\label{graveqsn}
We will begin our discussion of the physical structure of the theory~\eqref{neocon} by considering the Lagrangian field equations.
We borrow from~\cite{blagojevic2002gravitation} the definition of the generalised momenta
\begin{subequations}
\begin{align}
  \tensor{\pi}{_{i}^{kl}}&
    \equiv\frac{\partial bL_{\text{G}}}{\partial\tensor{\partial}{_\nu}\tensor{b}{^i_\mu}}\tensor{b}{^k_\mu}\tensor{b}{^l_\nu}
    \equiv-\frac{\partial bL_{\text{G}}}{\partial\tensor{T}{^{i}_{\mu\nu}}}\tensor{b}{^k_\mu}\tensor{b}{^l_\nu}
    \nonumber\\    
    &\equiv -2\planck^2b\sum_M\big(2\bet{M}\tensor{\mathcal{  T}}{^j_{nm}}+\cbet{M}\tensor{\lambda}{^j_{nm}}\big)\projlore[_j^{nm}_i^{kl}]{M},\label{equbeg}\\
  \tensor{\pi}{_{ij}^{kl}}&
  \equiv\frac{\partial bL_{\text{G}}}{\partial\tensor{\partial}{_\nu}\tensor{A}{^{ij}_\mu}}\tensor{b}{^k_\mu}\tensor{b}{^l_\nu}
    \equiv-\frac{\partial bL_{\text{G}}}{\partial\tensor{R}{^{ij}_{\mu\nu}}}\tensor{b}{^k_\mu}\tensor{b}{^l_\nu}
    \equiv 2\alp{0}\planck^2\tensor*{\delta}{_{[i}^k}\tensor*{\delta}{_{j]}^l}
    \nonumber\\    
    &\ \  -4b\sum_I\big(2\alp{I}\tensor{\mathcal{  R}}{^{pq}_{nm}}+\calp{I}\tensor{\lambda}{^{pq}_{nm}}\big)\projlore[_{pq}^{nm}_{ij}^{kl}]{I},\label{equbegtor}
\end{align}
\end{subequations}
where $b\equiv\det \tensor{b}{^i_\mu}$ plays the role of $\sqrt{-g}$ in GR, 
and we note for later convenience that, for \emph{any} values adopted by the various couplings, these quantities can be shown after a somewhat lengthy calculation to satisfy the identities
  \begin{align}
    \tensor{\mathcal{  T}}{_{[j|}_{pq}}\tensor{\pi}{_{|i]}^{pq}}
    -&2\tensor{\mathcal{  T}}{^p_{k[i|}}\tensor{\pi}{_p^{k}_{|j]}} 
    \nonumber\\
    &\equiv 
    \tensor{\mathcal{  R}}{^k_{[i|pq}}\tensor{\pi}{_{k|j]}^{pq}}
    +\tensor{\mathcal{  R}}{^{pq}_{k[i|}}\tensor{\pi}{_{pq}^k_{|j]}} \equiv 0.
    \label{convenience}
  \end{align}
  In terms of the generalised momenta we then obtain the stress-energy and spin field equations of the theory in the presence of matter sources, where $\tensor{D}{_\mu}\equiv\tensor{\partial}{_\mu}+\frac{1}{2}\tensor{A}{^{kl}_\mu}\tensor{\Sigma}{_{kl}}\cdot$ is the gauge-covariant derivative and $\tensor{\Sigma}{_{kl}}$ the (representation-specific) Lorentz group generators,
  \begin{subequations}
    \begin{align}
      \tensor{\tau}{^\nu_i}&=
      -\tensor{D}{_\mu}\tensor{\pi}{_i^{\nu\mu}}+\tensor{\mathcal{  T}}{^p_{ki}}\tensor{\pi}{_p^{k\nu}}+\frac{1}{2}\tensor{\mathcal{  R}}{^{pq}_{ki}}\tensor{\pi}{_{pq}^{k\nu}}
     +b\tensor{L}{_{\text{G}}}\tensor{h}{_i^\nu},
      \nonumber\\
      \tensor{\tau}{^\mu_\nu}&\equiv \tensor{h}{_k^\mu}\frac{\delta bL_{\text{M}}}{\delta \tensor{h}{_k^\nu}}\equiv -\frac{\delta bL_{\text{M}}}{\delta \tensor{b}{^k_\mu}}\tensor{b}{^k_\nu},\label{ergeqn}\\
    \tensor{\sigma}{^\nu_{ij}}&=
    -\tensor{D}{_\mu}\tensor{\pi}{_{ij}^{\nu\mu}} 
    +2\tensor{\pi}{_{[ij]}^\nu}
    \quad
     \tensor{\sigma}{^\mu_{ij}}\equiv-\frac{\delta bL_{\text{M}}}{\delta \tensor{A}{^{ij}_\mu}}.
     \label{spieqn}
  \end{align}
\end{subequations}
These equations may be manipulated further. We see that the divergence of the spin equation~\eqref{spieqn} is
\begin{equation}
  \tensor{D}{_\mu}\tensor{\sigma}{^\mu_{ij}}=
  2\tensor{D}{_\mu}\tensor{\pi}{_{[ij]}^\mu}
  +\tensor{\mathcal{  R}}{^k_{[i|pq}}\tensor{\pi}{_{k|j]}^{pq}},
  \label{spidiv}
\end{equation}
and we can expand the energy-momentum equation~\eqref{ergeqn} to give 
\begin{equation}
  \begin{aligned}
    \tensor{\tau}{^j_i}&=
    -\tensor{D}{_\mu}\tensor{\pi}{_i^{j\mu}}
    -\frac{1}{2}\tensor{\mathcal{  T}}{^j_{pq}}\tensor{\pi}{_i^{pq}}
    +\tensor{\mathcal{  T}}{^p_{ki}}\tensor{\pi}{_p^{kj}}
    \\
   &\ \ \ +\frac{1}{2}\tensor{\mathcal{  R}}{^{pq}_{ki}}\tensor{\pi}{_{pq}^{kj}}
    +b\tensor{L}{_{\text{G}}}\tensor*{\delta}{_i^j}.\label{ergexp}
  \end{aligned}
\end{equation}
However, by considering the skew-symmetric part of~\eqref{ergexp} and the conservation law $\tensor{D}{_\mu}\tensor{\sigma}{^\mu_{ij}}\equiv 2\tensor{\tau}{_{[ij]}}$, we see that there is another relation
\begin{equation}
  \tensor{D}{_\mu}\tensor{\sigma}{^\mu_{ij}}=
  2\tensor{D}{_\mu}\tensor{\pi}{_{[ij]}^\mu}
  -\tensor{\mathcal{  R}}{^{pq}_{k[i|}}\tensor{\pi}{_{pq}^k_{|j]}}.
  \label{anorel}
\end{equation}
From~\eqref{anorel} and~\eqref{spidiv} we can use the identities~\eqref{convenience} to confirm the gravitational equivalent of the conservation law, i.e. that six of the field equations are in fact shared between~\eqref{ergeqn} and~\eqref{spieqn}.
In the simple case of the teleparallel theory, we note that this result may be used to identify the so-called `$\lambda$ symmetry', i.e. the parts of $\tensor{\lambda}{_i^{jk}}$ which remain dynamically undetermined~\cite{2000CQGra..17.3785B}.

The most striking consequence of the geometric multipliers in~\eqref{neocon} follows from their own field equations, which suppress various parts of the Riemann--Cartan and torsion tensors. The first opportunity to employ the transfer couplings from~\cref{formalism} arises when we decompose these field equations into their respective $\othree$ irreps, to give
\begin{subequations}
  \begin{gather}
\begingroup 
\setlength\arraycolsep{0pt}
  \begin{pmatrix}
    \calpm[A]{\parallel\parallel} & \calpm[A]{\parallel\perp} \\
    \calpm[A]{\perp\parallel} & \calpm[A]{\perp\perp}
  \end{pmatrix}
  \endgroup
  \begin{pmatrix}
    \projorthhum[_{\acu{l}}_{nm}^{\ovl{pq}}]{A}\tensor{\mathcal{  R}}{^{nm}_{\ovl{pq}}} \\
    2\projorthhum[_{\acu{l}}_{nm}^{\ovl{q}}]{A}\tensor{\mathcal{  R}}{^{nm}_{\perp\ovl{q}}}
  \end{pmatrix}
  \approx
  \mathbf{0},\label{sysriex}\\
\begingroup 
\setlength\arraycolsep{0pt}
  \begin{pmatrix}
    \cbetm[E]{\parallel\parallel} & \cbetm[E]{\parallel\perp} \\
    \cbetm[E]{\perp\parallel} & \cbetm[E]{\perp\perp}
  \end{pmatrix}
  \endgroup
  \begin{pmatrix}
    \projorthhum[_{\acu{l}}_{n}^{\ovl{pq}}]{E}\tensor{\mathcal{  T}}{^{n}_{\ovl{pq}}} \\
    2\projorthhum[_{\acu{l}}_{n}^{\ovl{q}}]{E}\tensor{\mathcal{  T}}{^{n}_{\perp\ovl{q}}}
  \end{pmatrix}
  \approx
  \mathbf{0}.
  \label{sysrie}
\end{gather}
\end{subequations}
The consequences of the geometric multipliers are thus fully encoded by the pre-multiplying matrices in~\cref{sysrie,sysriex}.
Less formally, we provide in~\cref{mulcou} a translation of~\cref{sysrie,sysriex} in terms of the `human readable' $\othree$ representations of the Riemann--Cartan curvature and torsion.

Multipliers imposed to correct pathologies in the original PGT should not intefere with the desirable phenomenology, as established for example in~\cite{chapter2,chapter3}. This principle of \emph{selective non-interference} can be implemented by choosing the multiplier couplings so that
\begin{equation}
  \sum_{I}\calp{I}\projlore[_{ij}^{kl}_{pq}^{mn}]{I}\tensor{\mathcal{  R}}{^{pq}_{nm}}\approx
  \sum_{M}\cbet{M}\projlore[_{i}^{kl}_{p}^{mn}]{M}\tensor{\mathcal{  T}}{^{p}_{nm}}\approx 0,
  \label{selnon}
\end{equation}
on the phase-space shell defined by all desirable solutions to the original theory. These solutions are then \emph{still valid} for all multiplier extensions of the original theory which obey~\eqref{selnon}, so long as the multipliers themselves solve the coupled, homogeneous, first-order linear system
\begin{subequations}
\begin{gather}
      -\tensor{D}{_\mu}\tensor{\Lambda}{_i^{\nu\mu}}+\tensor{\mathcal{  T}}{^p_{ki}}\tensor{\Lambda}{_p^{k\nu}}
      +\tensor{\mathcal{  R}}{^{pq}_{ki}}\tensor{\Lambda}{_{pq}^{k\nu}}\approx 0,
      \\
    -\tensor{D}{_\mu}\tensor{\Lambda}{_{ij}^{\nu\mu}} 
    +\tensor{\Lambda}{_{[ij]}^\nu}\approx 0,\label{suplam}
\end{gather}
\end{subequations}
which is derived from~\cref{ergeqn,spieqn}, and expressed in terms of the `employed' multiplier d.o.f
\begin{subequations}
\begin{gather}
  \tensor{\Lambda}{_{ij}^{kl}}\equiv
  \sum_{I}\calp{I}\projlore[_{ij}^{kl}_{pq}^{mn}]{I}\tensor{\lambda}{^{pq}_{nm}},
  \label{mancorx}
  \\
  \tensor{\Lambda}{_{i}^{kl}}\equiv
  \planck^2\sum_{M}\cbet{M}\projlore[_{i}^{kl}_{p}^{mn}]{M}\tensor{\lambda}{^{p}_{nm}}.
  \label{mancor}
\end{gather}
\end{subequations}
Formally, the system~\eqref{suplam} can always be satisfied (e.g. with vanishing multipliers), though attention must still be paid to the uniqueness of such solutions for a given spacetime symmetry, along with the physical interpretation of the multipliers.

\section{The Hamiltonian picture}\label{hampiq}
Having briefly examined the Lagrangian formulation of geometric multipliers in~\cref{lagpic}, we now turn to the Hamiltonian formulation. The Hamiltonian structure of the conventional PGT is well understood, and presented clearly in~\cite{blagojevic2002gravitation}. We use the same conventions as in~\cite{blagojevic2002gravitation,chapter4,mythesis}.
\subsection{The new super-Hamiltonian}\label{newsuperham}
The total Hamiltonian $\tensor{\mathcal{  H}}{_{\text{T}}}$ for the PGT, as we have written it in~\cite{chapter4}, is extended by the geometric multipliers to
\begin{equation}
  \begin{aligned}
    \tensor{\mathcal{  H}}{_{\text{T}}}&\equiv\tensor{\mathcal{  H}}{_{\text{C}}}+\tensor{u}{^k_0}\tensor{\varphi}{_k^0}+\frac{1}{2}\tensor{u}{^{jk}_0}\tensor{\varphi}{_{jk}^0}+(u\cdot\varphi)
  \\
  &\ \ \ +\tensor{\upsilon}{_{i}^{kl}}\tensor{\phi}{^{i}_{kl}}
  +\tensor{\upsilon}{_{ij}^{kl}}\tensor{\phi}{^{ij}_{kl}}.
  \label{tottr}
  \end{aligned}
\end{equation}
We shall now account for all the quantities appearing here.
The \emph{canonical} Hamiltonian $\tensor{\mathcal{H}}{_{\text{C}}}$ is formed by Legendre transforming the Lagrangian over the \emph{canonical} momenta, the timelike parts of~\cref{equbeg,equbegtor}
\begin{equation}
  \tensor{\pi}{_i^\mu}\equiv\frac{\partial bL_{\text{G}}}{\partial(\partial_0\tensor{b}{^i_{\mu}})}, \quad \tensor{\pi}{_{ij}^{\mu}}\equiv\frac{\partial bL_{\text{G}}}{\partial(\partial_0\tensor{A}{^{ij}_{\mu}})}.
  \label{canonicalmomenta}
\end{equation}
In the \emph{Dirac} form~\cite{1987PhRvD..35.3748B,PhysRevD.30.2508}, this quantity is
\begin{equation}
  \tensor{\mathcal{H}}{_{\text{C}}}\equiv N\tensor{\mathcal{H}}{_\perp}+\tensor{N}{^\alpha}\tensor{\mathcal{H}}{_\alpha}-\frac{1}{2}\tensor{A}{^{ij}_0}\tensor{\mathcal{H}}{_{ij}}+\tensor{\partial}{_\alpha}\tensor{\mathscr{D}}{^\alpha},
  \label{canonicalhamiltonian}
\end{equation}
i.e. linear in the non-physical \emph{lapse} function and \emph{shift} vector, which are defined with reference to the non-physical part of the translational gauge field $N\equiv\tensor{n}{_k}\tensor{b}{^k_0}$, and $\tensor{N}{^\alpha}\equiv\tensor{h}{_{\ovl{k}}^\alpha}\tensor{b}{^{\ovl{k}}_0}$. 
The functions in~\eqref{canonicalhamiltonian} are 
\begin{subequations}
\begin{align}
  \tensor{\mathcal{H}}{_\perp} &\equiv\tensor{\hat{\pi}}{_i^{\overline{k}}}\tensor{\mathcal{  T}}{^i_{\perp\overline{k}}}+\frac{1}{2}\tensor{\hat{\pi}}{_{ij}^{\overline{k}}}\tensor{\mathcal{  R}}{^{ij}_{\perp\overline{k}}}-JL_{\text{G}}-\tensor{n}{^k}\tensor{D}{_\alpha}\tensor{\pi}{_k^\alpha},\label{total_hamiltonian}\\
  \tensor{\mathcal{H}}{_\alpha} & \equiv\tensor{\pi}{_i^\beta}\tensor{T}{^i_{\alpha\beta}}+\frac{1}{2}\tensor{\pi}{_{ij}^\beta}\tensor{R}{^{ij}_{\alpha\beta}}-\tensor{b}{^k_\alpha}\tensor{D}{_\beta}\tensor{\pi}{_k^\beta},\label{total_hamiltonian_start}\\
  \tensor{\mathcal{H}}{_{ij}}& \equiv 2\tensor{\pi}{_{[i}^\alpha}\tensor{b}{_{j]\alpha}}+\tensor{D}{_\alpha}\tensor{\pi}{_{ij}^\alpha},\label{total_hamiltonian_int}\\
  \tensor{\mathscr{D}}{^\alpha}& \equiv \tensor{b}{^i_0}\tensor{\pi}{_i^\alpha}+\frac{1}{2}\tensor{A}{^{ij}_0}\tensor{\pi}{_{ij}^\alpha},\label{total_hamiltonian_fin}
\end{align}
\end{subequations}
where the `parallel' momenta are $\smash{\tensor{\hat{\pi}}{_{i}^{\overline{k}}}\equiv\tensor{\pi}{_{i}^\alpha}\tensor{b}{^{k}_\alpha}}$ and $\smash{\tensor{\hat{\pi}}{_{ij}^{\overline{k}}}\equiv\tensor{\pi}{_{ij}^\alpha}\tensor{b}{^{k}_\alpha}}$.
The field strengths defined in~\eqref{riemanndef} and~\eqref{torsiondef}, and which appear in~\eqref{neocon}, are independent of the velocities of $\tensor{b}{^k_0}$ and $\tensor{A}{^{ij}_0}$. It follows that the theory has (at least) the $10$ primary constraints
\begin{equation}
  \tensor{\varphi}{_k^0}\equiv\tensor{\pi}{_k^0}\approx 0, \quad \tensor{\varphi}{_{ij}^0}\equiv\tensor{\pi}{_{ij}^0}\approx 0,
  \label{sureprimaries}
\end{equation}
where ($\approx$) denotes weak equality on the phase shell, and from~\eqref{canonicalhamiltonian} we see that the consistency of~\eqref{sureprimaries} invokes the `sure' \emph{secondary} first class (sSFC) constriants
\begin{equation}
  \tensor{\mathcal{H}}{_\perp}\approx 0, \quad \tensor{\mathcal{H}}{_\alpha}\approx 0, \quad \tensor{\mathcal{H}}{_{ij}}\approx 0.
  \label{suresecondaries}
\end{equation}
Depending on the parameters $\{\alp{I}\}$, $\{\bet{M}\}$, there may be other primary \emph{if} constraints (PiCs), denoted in~\eqref{tottr} by
\begin{equation}
  \begin{aligned}
  (u\cdot\varphi)&\equiv
  \frac{1}{32}\sum_{A}
  \ctmp[A]{\perp}\nu(\alpm[A]{\perp\perp})
  \tensor[^A]{u}{_{\acu{v}}}\tensor[^A]{\varphi}{^{\acu{v}}}
  \\
  &\ \ \ +\frac{1}{8\planck}\sum_{E}
  \ctmp[E]{\perp}\nu(\betm[E]{\perp\perp})
  \tensor[^E]{u}{_{\acu{v}}}\tensor[^E]{\varphi}{^{\acu{v}}}
  .\label{new_pics}
  \end{aligned}
\end{equation}
We note a change in~\eqref{new_pics} from the previous formalism in~\cite{chapter4}, in that we introduce factors of $\ctmp[A]{\perp}/32$ and $\ctmp[E]{\perp}/8$ -- this just amounts to a rescaling of the Hamiltonian multipliers\footnote{Care should be taken to distinguish between \emph{Hamiltonian} and \emph{geometric} (i.e. \emph{Lagrangian}) multipliers.} $\tensor[^A]{u}{^{\acu{v}}}$ by some constants at the point of definition, and will make things more convenient in~\cref{conpri}. The PiC functions are now
\begin{subequations}
  \begin{align}
  \tensor[^A]{\varphi}{_{\acu{v}}}&\equiv\frac{1}{J}\tensor[^A]{\hat{\pi}}{_{\acu{v}}}
  +2\alp{0}\planck^2\projorthhum[_{\acu{v}\perp\ovl{k}}^{\ovl{k}}]{A}
  -8\calpm[A]{\perp\perp}\projorthhum[_{\acu{v}}_{jk}^{\ovl{m}}]{A}\tensor{\lambda}{^{jk}_{\perp\ovl{m}}}
  \nonumber\\
  &\ \ \ -4\projorthhum[_{\acu{v}}_{jk}^{\ovl{lm}}]{A} 
  \Big(\calpm[A]{\perp\parallel}\tensor{\lambda}{^{jk}_{\ovl{lm}}}+2\alpm[A]{\perp\parallel}\tensor{\mathcal{  R}}{^{jk}_{\ovl{lm}}}\Big),
  \label{fulpic}\\
  \tensor[^E]{\varphi}{_{\acu{v}}}&\equiv\frac{1}{J}\tensor[^E]{\hat{\pi}}{_{\acu{v}}}
  -4\planck^2\cbetm[E]{\perp\perp}\projorthhum[_{\acu{v}}_{j}^{\ovl{m}}]{E}\tensor{\lambda}{^{j}_{\perp\ovl{m}}}
  \nonumber\\
  &\ \ \ -2\planck^2\projorthhum[_{\acu{v}}_{j}^{\ovl{lm}}]{E} 
  \Big(\cbetm[E]{\perp\parallel}\tensor{\lambda}{^{j}_{\ovl{lm}}}+2\betm[E]{\perp\parallel}\tensor{\mathcal{  T}}{^{j}_{\ovl{lm}}}\Big),
  \label{fulpictor}
\end{align}
\end{subequations}
so they generally acquire a dependency on the multiplier fields.
Note that we use the foliation measure $J\equiv b/N$ .
Within the canonical Hamiltonian defined in~\eqref{canonicalhamiltonian}, the super-Hamiltonian in~\eqref{total_hamiltonian} is modified beyond the formula in~\cite{chapter4} to
\begin{equation}
\begin{aligned}
  \tensor{\mathcal{H}}{_\perp}&\equiv\frac{J}{64}\sum_{A}\ctmp[A]{\perp}\mu(\alpm[A]{\perp\perp})\tensor[^A]{\varphi}{_{\acu{v}}}\tensor[^A]{\varphi}{^{\acu{v}}}\\
  &\ \ \ +\frac{J}{16\planck}\sum_{E}\ctmp[E]{\perp}\mu(\betm[E]{\perp\perp})\tensor[^E]{\varphi}{_{\acu{v}}}\tensor[^E]{\varphi}{^{\acu{v}}}+\frac{1}{2}\alp{0}\planck^2\cR[]{A0p}\\
    &\ \ \ -J\sum_{I}\Big(\alp{I}\tensor{\mathcal{  R}}{^{ij}_{\ovl{kl}}}+\calp{I}\tensor{\lambda}{^{ij}_{kl}}\Big)\projlore[_{ij}^{kl}_{nm}^{\ovl{pq}}]{I}\tensor{\mathcal{  R}}{^{nm}_{\ovl{pq}}}\\
    &\ \ \ -J\planck^2\sum_{M}\Big(\bet{M}\tensor{\mathcal{  T}}{^{i}_{\ovl{kl}}}+\cbet{M}\tensor{\lambda}{^{i}_{kl}}\Big)\projlore[_{i}^{kl}_{n}^{\ovl{pq}}]{M}\tensor{\mathcal{  T}}{^{n}_{\ovl{pq}}}\\
    &\ \ \ -\tensor{n}{^k}\tensor{D}{_\alpha}\tensor{\pi}{_k^\alpha}.
    \label{ghoul}
\end{aligned}
\end{equation}
The remaining parts, the linear and rotational supermomenta~\cref{total_hamiltonian_start,total_hamiltonian_int} and the surface term~\cref{total_hamiltonian_fin}, are as defined in~\cite{chapter4}.
Finally, the primary constraints
\begin{equation}
  \tensor{\phi}{^{ij}_{kl}}\equiv \tensor{\varpi}{^{ij}_{kl}}\approx 0,\quad
\tensor{\phi}{^{i}_{kl}}\equiv \tensor{\varpi}{^{i}_{kl}}\approx 0,
  \label{newp}
\end{equation}
i.e. the naturally defined \emph{multiplier momenta}
\begin{equation}
  \tensor{\varpi}{_i^{jk}}\equiv\frac{\partial bL_{\text{G}}}{\partial(\partial_0\tensor{\lambda}{^i_{jk}})}, \quad \tensor{\varpi}{_{ij}^{kl}}\equiv\frac{\partial bL_{\text{G}}}{\partial(\partial_0\tensor{\lambda}{^{ij}_{kl}})},
  \label{multipliermomenta}
\end{equation}
must be introduced because the Lagrangian~\eqref{neocon} is independent of the multiplier velocities $\tensor{\dot\lambda}{^{ij}_{kl}}$ and $\tensor{\dot\lambda}{^{i}_{kl}}$. 
\subsection{Consistency of geometric primaries}\label{conpri}
We will now consider the application of the Dirac--Bergmann algorithm on the general theory, and discover a substantial departure from the simple teleparallel constraint structure of~\eqref{fullteleparallel}.
In what follows we will discuss the effects of Riemann--Cartan and torsion multipliers concurrently; while these sectors differ in certain numerical factors and notation, the discussion is essentially the same up to some placeholder results in the torsion sector.
We first see that there is a new pair of secondary constraints from~\eqref{newp}, $\tensor{\chi}{^{ij}_{kl}}\equiv\tensor{\dot{\varpi}}{^{ij}_{kl}}\approx 0$ and $\tensor{\chi}{^{i}_{kl}}\equiv\tensor{\dot{\varpi}}{^{i}_{kl}}\approx 0$, which we find to be equivalent to
  \begin{subequations}
  \begin{align}
    \sum_{I}&\calp{I}\projlore[_{ij}^{kl}_{nm}^{pq}]{I}\bigg[     
    8b\tensor{\mathcal{  R}}{^{nm}_{\ovl{pq}}}
    \nonumber\\
    +\sum_{A}&
    \ctmp[A]{\perp}\Big(
	b\mu(\alpm[A]{\perp\perp})\tensor[^A]{\varphi}{^{\acu{r}}}
      +\nu(\alpm[A]{\perp\perp})\tensor[^A]{u}{^{\acu{r}}}
    \Big)\foli{p}\projorthhum[_{\acu{r}}^{nm}_{\ovl{q}}]{A}
    \bigg]
    \approx 0,
    \label{skilor}\\
    \sum_{M}&\cbet{M}\projlore[_{i}^{kl}_{n}^{pq}]{M}\bigg[     
    4\planck^2b\tensor{\mathcal{  T}}{^{n}_{\ovl{pq}}}
    \nonumber\\
    +\sum_{E}&
    \ctmp[E]{\perp}\Big(
	b\mu(\betm[E]{\perp\perp})\tensor[^E]{\varphi}{^{\acu{r}}}
      +\nu(\betm[E]{\perp\perp})\tensor[^E]{u}{^{\acu{r}}}
    \Big)\foli{p}\projorthhum[_{\acu{r}}^{n}_{\ovl{q}}]{E}
    \bigg]
      \approx 0.
    \label{skilup}
  \end{align}
\end{subequations}
These secondaries correspond to two statements in the teleparallel theory, eliminating the Riemann--Cartan curvature and a multiplier. In the general theory we obtain by projections of~\cref{skilor,skilup} multiple possible statements which can be written more compactly as 
\begin{subequations}
  \begin{gather}
\begingroup 
\setlength\arraycolsep{0pt}
  \begin{pmatrix}
    \calpm[A]{\parallel\parallel} & \calpm[A]{\parallel\perp} \\
    \calpm[A]{\perp\parallel} & \calpm[A]{\perp\perp}
  \end{pmatrix}
  \begin{pmatrix}
    8b\projorthhum[_{\acu{l}}_{nm}^{\ovl{pq}}]{A}\tensor{\mathcal{  R}}{^{nm}_{\ovl{pq}}} \\
    b\mu(\alpm[A]{\perp\perp})\tensor[^A]{\varphi}{_{\acu{l}}}+\nu(\alpm[A]{\perp\perp})\tensor[^A]{u}{_{\acu{l}}}
  \end{pmatrix}
  \endgroup
  \approx
  \mathbf{0},\label{systemx}
  \\
\begingroup 
\setlength\arraycolsep{0pt}
  \begin{pmatrix}
    \cbetm[E]{\parallel\parallel} & \cbetm[E]{\parallel\perp} \\
    \cbetm[E]{\perp\parallel} & \cbetm[E]{\perp\perp}
  \end{pmatrix}
  \endgroup
  \begin{pmatrix}
    4\planck^2b\projorthhum[_{\acu{l}}_{n}^{\ovl{pq}}]{E}\tensor{\mathcal{  T}}{^{n}_{\ovl{pq}}} \\
    b\mu(\betm[E]{\perp\perp})\tensor[^E]{\varphi}{_{\acu{l}}}+\nu(\betm[E]{\perp\perp})\tensor[^E]{u}{_{\acu{l}}}
  \end{pmatrix}
  \approx
  \mathbf{0}.
\label{system}
\end{gather}
\end{subequations}
We note from~\cref{system,systemx} the counterpart in the Hamiltonian picture of the linear systems first encountered in~\cref{sysrie,sysriex}.
For any of the sectors $A$ and $E$, we will next discuss the implications of these systems for various $\{\calp{I}\}$ and $\{\cbet{M}\}$, and then in each sub-case for various $\{\alp{I}\}$ and $\{\bet{M}\}$.
\paragraph{Sector is not multiplier-constrained}
Sometimes we will have results such as
\begin{subequations}
  \begin{gather}
    \calpm[A]{\parallel\parallel}=\calpm[A]{\perp\perp}=\calpm[A]{\parallel\perp}=\calpm[A]{\perp\parallel}=0,\label{unconx}\\
  \cbetm[E]{\parallel\parallel}=\cbetm[E]{\perp\perp}=\cbetm[E]{\parallel\perp}=\cbetm[E]{\perp\parallel}=0,
  \label{uncon}
\end{gather}
\end{subequations}
in which case we proceed normally, as we did in~\cite{chapter4} -- i.e. we follow the conventional \emph{if-constraint} formalism set out in~\cite{1983PhRvD..28.2455B,1987PhRvD..35.3748B}.
\paragraph{Sector is multiplier-constrained and non-singular}
More generally we will have 
  \begin{equation}
  \calpm[A]{\parallel\parallel}\calpm[A]{\perp\perp}-\calpm[A]{\parallel\perp}\calpm[A]{\perp\parallel}\neq 0, \quad
  \cbetm[E]{\parallel\parallel}\cbetm[E]{\perp\perp}-\cbetm[E]{\parallel\perp}\cbetm[E]{\perp\parallel}\neq 0,
  \label{nonsingular}
\end{equation}
but not~\cref{uncon,unconx}, in which case one or both of the systems~\cref{system,systemx} is only satisfied by a vanishing vector. In this case the first vanishing component of either system \emph{always} gives us a secondary constraint
\begin{subequations}
\begin{gather}
  \tensor*[^A]{\chi}{^{\parallel}_{\acu{v}}}\equiv
  \projorthhum[_{\acu{v}}_{nm}^{\ovl{pq}}]{A}\tensor{\mathcal{  R}}{^{nm}_{\ovl{pq}}}\approx 0, \label{canconx}\\
  \tensor*[^E]{\chi}{^{\parallel}_{\acu{v}}}\equiv
  \planck^2\projorthhum[_{\acu{v}}_{n}^{\ovl{pq}}]{E}\tensor{\mathcal{  T}}{^{n}_{\ovl{pq}}}\approx 0,
  \label{cancon}
\end{gather}
\end{subequations}
independently of the $\{\alp{I}\}$ or $\{\bet{M}\}$. The parallel parts of the field strengths can then be safely eliminated from the corresponding PiC functions in~\cref{fulpic,fulpictor}.
However, we recall that this PiC function is only a PiC if ${\nu(\alpm[A]{\perp\perp})=1}$ or ${\nu(\betm[E]{\perp\perp})=1}$. In that case, the second vanishing component does not give us a further secondary constraint, but instead determines a multiplier
\begin{equation}
  \tensor[^A]{u}{_{\acu{v}}}\approx 0, \quad
  \tensor[^E]{u}{_{\acu{v}}}\approx 0,
  \label{mulcon}
\end{equation}
and so we see that the PiC associated with sector $A$ or $E$ spontaneously becomes second class (SC). If on the other hand we have ${\nu(\alpm[A]{\perp\perp})=0}$ or ${\nu(\betm[E]{\perp\perp})=0}$, then the vanishing of the second component means that the PiC function (with its field strength terms removed) becomes a further secondary constraint, 
\begin{equation}
  \tensor*[^A]{\chi}{^{\perp}_{\acu{v}}}\equiv\tensor[^A]{\varphi}{_{\acu{v}}}\approx 0, \quad
  \tensor*[^E]{\chi}{^{\perp}_{\acu{v}}}\equiv\tensor[^E]{\varphi}{_{\acu{v}}}\approx 0,
  \label{fursec}
\end{equation}
even though it was not primarily constrained by the $\{\alp{I}\}$ or $\{\bet{M}\}$.

We are now in a position to confirm the action of the Lorentz constraint in the Lagrangian picture. We see from the Hamiltonian e.o.m that
  \begin{subequations}
  \begin{align}
    \tensor{\dot{A}}{^{ij}_{\alpha}}\equiv
    \tensor{\partial}{_\alpha}\tensor{A}{^{ij}_{0}}
    +2\tensor{A}{^{l[j}_0}\tensor{A}{_l^{i]}_{\alpha}}
    +&\tensor{N}{^\beta}\tensor{R}{^{ij}_{\beta\alpha}}
    \nonumber\\
    +\frac{1}{64}\frac{\partial}{\partial\tensor{\pi}{_{ij}^{\alpha}}}
    \sum_{A}\ctmp[A]{\perp}\Big(&
	b\mu(\alpm[A]{\perp\perp})\tensor[^A]{\varphi}{_{\acu{v}}}
	\nonumber\\
	&
	+2\nu(\alpm[A]{\perp\perp})\tensor[^A]{u}{_{\acu{v}}}
    \Big)
    \tensor[^A]{\varphi}{^{\acu{v}}}
    ,\\
    \tensor{\dot{b}}{^{i}_{\alpha}}\equiv
    \tensor{\partial}{_\alpha}\tensor{b}{^{i}_{0}}
    +\tensor{b}{^l_{0}}\tensor{A}{^i_{l\alpha}}
    -\tensor{A}{^i_{j0}}&\tensor{b}{^j_{\alpha}}
    +\tensor{N}{^\beta}\tensor{T}{^{i}_{\beta\alpha}}
    \nonumber\\
    +\frac{1}{16\planck}\frac{\partial}{\partial\tensor{\pi}{_{i}^{\alpha}}}
    \sum_{E}\ctmp[E]{\perp}\Big(&
	b\mu(\betm[E]{\perp\perp})\tensor[^E]{\varphi}{_{\acu{v}}}
	\nonumber\\
	&
	+2\nu(\betm[E]{\perp\perp})\tensor[^E]{u}{_{\acu{v}}}
    \Big)
    \tensor[^E]{\varphi}{^{\acu{v}}}
    ,
  \end{align}
\end{subequations}
and by rearranging this and projecting, we find a useful general expression for the \emph{velocity} parts of the Riemann--Cartan and torsion tensors in terms of canonical quantities
  \begin{subequations}
\begin{gather}
    16b\projorthhum[_{\acu{v}}_{ij}^{\ovl{k}}]{A}\tensor{\mathcal{  R}}{^{ij}_{\perp\ovl{k}}}\equiv 
    b\mu(\alpm[A]{\perp\perp})\tensor[^A]{\varphi}{_{\acu{v}}}
    +\nu(\alpm[A]{\perp\perp})\tensor[^A]{u}{_{\acu{v}}},
    \label{velrelx}
    \\
    8b\projorthhum[_{\acu{v}}_{i}^{\ovl{k}}]{E}\tensor{\mathcal{  T}}{^{i}_{\perp\ovl{k}}}\equiv 
    b\mu(\betm[E]{\perp\perp})\tensor[^E]{\varphi}{_{\acu{v}}}
    +\nu(\betm[E]{\perp\perp})\tensor[^E]{u}{_{\acu{v}}}.
  \label{velrel}
\end{gather}
\end{subequations}
Recall that these velocities are not part of the constraint algebra, and are usually found to be multipliers -- we could have used this expression for example in~\cite{chapter4}. However it is now clear from~\cref{velrel,velrelx} and from~\eqref{mulcon} and~\eqref{fursec} that when the sector $A$ or $E$ is multiplier-constrained and non-singular, the velocity parts of the Riemann--Cartan or torsion tensors in that sector will be vanishing, no matter what is $\mu(\alpm[A]{\perp\perp})$ or $\mu(\betm[E]{\perp\perp})$. In combination with the canonical constraint~\cref{cancon,canconx}, this means that the \emph{whole} of the field strength tensor in the $A$ or $E$ sector vanishes, which is precisely the effect in~\cref{sysrie,sysriex} of the multipliers in the Lagrangian picture.
\paragraph{Sector is multiplier-constrained and singular}
There is a special case where neither~\cref{uncon,unconx} nor~\eqref{nonsingular} are true. When one of the matrices is singular in this way, both consistency conditions for each $A$ or $E$ are equivalent but nontrivial. Once again the outcome depends on the $\{\alp{I}\}$ and $\{\bet{M}\}$. If ${\nu(\alpm[A]{\perp\perp})=1}$ or ${\nu(\betm[E]{\perp\perp})=1}$, the original PiC function is indeed a PiC and
\begin{subequations}
  \begin{gather}
    \tensor[^A]{u}{_{\acu{v}}}\approx -\frac{8\calpm[A]{\perp\parallel}}{\calpm[A]{\perp\perp}}b\projorthhum[_{\acu{v}}_{nm}^{\ovl{pq}}]{A}\tensor{\mathcal{  R}}{^{nm}_{\ovl{pq}}}, \label{mulnewx}\\
  \tensor[^E]{u}{_{\acu{v}}}\approx -\frac{4\cbetm[E]{\perp\parallel}}{\cbetm[E]{\perp\perp}}\planck^2b\projorthhum[_{\acu{v}}_{n}^{\ovl{pq}}]{E}\tensor{\mathcal{  T}}{^{n}_{\ovl{pq}}},
  \label{mulnew}
\end{gather}
\end{subequations}
so the PiC is again SC. In this case no new secondaries are introduced.
Otherwise if ${\nu(\alpm[A]{\perp\perp})=0}$ or ${\nu(\betm[E]{\perp\perp})=0}$, a new secondary is introduced
\begin{subequations}
  \begin{gather}
  \tensor*[^A]{\chi}{^{\vDash}_{\acu{v}}}\equiv 
\tensor[^A]{\varphi}{_{\acu{v}}}
 +\frac{8\calpm[A]{\perp\parallel}\alpm[A]{\perp\perp}}{\calpm[A]{\perp\perp}}\projorthhum[_{\acu{v}}_{jk}^{\ovl{lm}}]{A}
  \tensor{\mathcal{  R}}{^{jk}_{\ovl{lm}}}
  \approx 0,
  \label{indcalx}
  \\
  \tensor*[^E]{\chi}{^{\vDash}_{\acu{v}}}\equiv 
\tensor[^E]{\varphi}{_{\acu{v}}}
 +\frac{4\cbetm[E]{\perp\parallel}\betm[E]{\perp\perp}}{\cbetm[E]{\perp\perp}}\planck^2\projorthhum[_{\acu{v}}_{j}^{\ovl{lm}}]{E}
  \tensor{\mathcal{  T}}{^{j}_{\ovl{lm}}}
  \approx 0.
  \label{indcal}
\end{gather}
\end{subequations}
Now again it is necessary to check the constraints from the Lagrangian picture. We see immediately from~\cref{velrel,velrelx,mulnew,mulnewx,indcal,indcalx} that the only such relations are
  \begin{subequations}
    \begin{gather}
    \calpm[A]{\parallel\parallel}\projorthhum[_{\acu{v}}_{nm}^{\ovl{pq}}]{A}\tensor{\mathcal{  R}}{^{nm}_{\ovl{pq}}}+2\calpm[A]{\parallel\perp}\projorthhum[_{\acu{v}}_{nm}^{\ovl{q}}]{A}\tensor{\mathcal{  R}}{^{nm}_{\perp\ovl{q}}}\approx 0,
    \\
    \cbetm[E]{\parallel\parallel}\projorthhum[_{\acu{v}}_{n}^{\ovl{pq}}]{E}\tensor{\mathcal{  T}}{^{n}_{\ovl{pq}}}+2\cbetm[E]{\parallel\perp}\projorthhum[_{\acu{v}}_{n}^{\ovl{q}}]{E}\tensor{\mathcal{  T}}{^{n}_{\perp\ovl{q}}}\approx 0.
  \end{gather}
  \end{subequations}
  Again, this is exactly what we expected for the singular case of~\cref{sysrie,sysriex}.
\subsection{Consistency of geometric secondaries}\label{consec}

In the canonical analysis of our new general theory~\eqref{neocon} we observe that the gravitational gauge fields introduce ${2\times ( 16+24 )}$ canonical d.o.f, and likewise ${2\times (24+36)}$ d.o.f are introduced by the geometric multipliers, for a total of 200 canonical d.o.f divided over 100 fields and 100 field momenta. 
Typically, $2\times m$ d.o.f will have been introduced formally through $m$ field d.o.fs allocated to `unemployed' multiplier irreps (unemployed in the sense of~\cref{mancor,mancorx}). Their elimination from the final counting is equally formal, since the corresponding $\soonethree$ irreps of their momenta (the primaries $\tensor{\varphi}{^{ij}_{kl}}$ and $\tensor{\varphi}{^{i}_{kl}}$ in~\eqref{newp}) will be first class (FC). 
The primarily constrained momenta of `employed' irreps are not obviously FC, since they fail to commute with their own secondaries as follows
\begin{equation}
\begin{aligned}
    \Big\{\tensor{\phi}{^{ij}_{kl}},\tensor*[^A]{\chi}{^{\perp}_{\acu{v}}}\Big\}&\approx \Big\{ \tensor{\phi}{^{ij}_{kl}},\tensor*[^A]{\chi}{^{\vDash}_{\acu{v}}}\Big\}
    \\
    \approx 16\Big(&\calpm[A]{\perp\parallel}\projorthhum[{_{\acu{v}}^{ij}_{\ovl{kl}}}]{A}
+2\calpm[A]{\perp\perp}\foli{[k|}\projorthhum[{_{\acu{v}}^{ij}_{|\ovl{l}]}}]{A}\Big)\delta^3,
  \\
  \Big\{\tensor{\phi}{^{i}_{kl}},\tensor*[^E]{\chi}{^{\perp}_{\acu{v}}}\Big\}&\approx \Big\{ \tensor{\phi}{^{i}_{kl}},\tensor*[^E]{\chi}{^{\vDash}_{\acu{v}}}\Big\}
  \\
  \approx 4\Big(&\cbetm[E]{\perp\parallel}\projorthhum[{_{\acu{v}}^{i}_{\ovl{kl}}}]{E}
  +2\cbetm[E]{\perp\perp}\foli{[k|}\projorthhum[{_{\acu{v}}^{i}_{|\ovl{l}]}}]{E}\Big)\delta^3.
\end{aligned}
\label{lincommal}
\end{equation}
We note however that every $J^P$ contains two momentum irreps, up to placeholder cases in the torsion sector, and from these parts we see that the combinations
\begin{subequations}
\begin{gather}
  2\ctmp[A]{\perp}\calpm[A]{\perp\perp}\projorthhum[{_{\acu{u}}_{ij}^{\ovl{kl}}}]{A}\tensor{\varpi}{^{ij}_{\ovl{kl}}}-\ctmp[A]{\parallel}\calpm[A]{\perp\parallel}\projorthhum[{_{\acu{u}}_{ij}^{\ovl{l}}}]{A}\tensor{\varpi}{^{ij}_{\perp\ovl{l}}}\approx0,
  \\
  2\ctmp[E]{\perp}\cbetm[E]{\perp\perp}\projorthhum[{_{\acu{u}}_{i}^{\ovl{kl}}}]{E}\tensor{\varpi}{^{i}_{\ovl{kl}}}-\ctmp[E]{\parallel}\cbetm[E]{\perp\parallel}\projorthhum[{_{\acu{u}}_{i}^{\ovl{l}}}]{E}\tensor{\varpi}{^{i}_{\perp\ovl{l}}}\approx0,
  \label{lot}
\end{gather}
\end{subequations}
commute with $\tensor*[^A]{\chi}{^{\perp}_{\acu{v}}}$ and $\tensor*[^A]{\chi}{^{\vDash}_{\acu{v}}}$ or $\tensor*[^E]{\chi}{^{\perp}_{\acu{v}}}$ and $\tensor*[^E]{\chi}{^{\vDash}_{\acu{v}}}$, and are in fact FC. We will not attempt here a general theory of the remaining consistency conditions. For such a theory, the effects of the $\{\calp{I}\}$, $\{\cbet{M}\}$ must in some sense be `multiplied' by those of the $\{\alp{I}\}$, $\{\bet{M}\}$, and the interactions are not obvious. For our purposes therefore, the consistencies of the remaining constraints must be obtained on a case-by-case basis.

We also recall that we must always subtract $2\times 10$ d.o.f due to the sPFCs, and in the nonlinear theory we tentatively assume all the $2\times 10$ sSFCs will be independent and must also be removed. We show separately in~\cref{linearisation_doubts,simpleA1p} how the sSFCs may be reduced or become degenerate in the linearised theory, if the Einstein--Hilbert term is absent.

Independently of their utility, the new commutators tend to suffer from an old challenge (noticed for example in~\cite{1983PhRvD..28.2455B}) as follows. 
While the parallel field strengths do express the fields conjugate to the field momenta, they also contain spatial gradients of those fields. 
Within the formal definition of the Poisson bracket, this can lead to gradients of the equal-time Dirac function, and an apparent loss of explicit covariance for the more complex expressions.
In~\cref{surfic} we clarify such situations by constructing a general and covariant expression for the Poisson bracket, which then takes the form of a differential operator.

\section{Minimal spin-parity $1^+$ theory}\label{lorspe}

Having introduced both the Lagrangian and Hamiltonian formulations of geometric multipliers, we now provide a brief illustration of how they might be used to combat the strong coupling problem in PGT. We use for our example the first of the `disallowed' PGTs from~\cite{2002IJMPD..11..747Y}, the $1^+$ theory which builds on the groundwork laid by Sezgin and van Nieuwenhuizen~\cite{1980PhRvD..21.3269S}. This theory is reached by imposing on~\eqref{neocon} the conditions
\begin{equation}
  \begin{aligned}
  \alp{1}&=\alp{2}=\alp{3}=\alp{4}=\alp{6}=\bet{1}=\bet{2}\\
  &=\calp{1}=\calp{2}=\calp{3}=\calp{4}=\calp{5}=\calp{6}\\
  &=\cbet{1}=\cbet{2}=\cbet{3}=0.
  \label{simspione}
\end{aligned}
\end{equation}
Note that the multiplier couplings, which are new in this work, must all be disabled. We provide in~\cref{simpleA1p} our attempt at the Hamiltonian analysis of this conventional theory. Our findings corroborate those in~\cite{2002IJMPD..11..747Y}, viz given certain assumptions (e.g. the validity of breaking consistency conditions over different $J^P$ pairs), a total of eight d.o.f seem to be propagating in the nonlinear theory. The extra three d.o.f are assumed to be a strongly coupled $1^-$ particle, since the PiC function $\pic[\ovl{i}]{A1m}$ is never constrained.

How to suppress the $1^-$ mode? By an examination of~\eqref{courel} and~\eqref{simspione} -- relations which are obeyed equally by the $\{\alp{I}\}$ -- we see that by setting $\alp{5}=0$ we can fix $\pic[\ovl{i}]{A1m}\approx 0$ and so constrain the momentum $\PiP[\ovl{i}]{A1m}$. However by doing so we will also fix $\pic[\ovl{ij}]{A1p}\approx 0$, and so deactivate the $1^+$ mode with which we started. We would prefer not to modify the Sezgin--van Nieuwenhuizen conditions at all, and so we instead try to introduce geometric multipliers. 

The natural multiplier approach will be to constrain $\PiP[\ovl{i}]{A1m}$ with $\calp{4}\neq 0$. In fact, this route turns out to become complicated due to the appearance of a singular secondary, and collateral effects in the $0^+$ and $2^+$ sectors. 
In general, we feel that rotational multipliers will be more dangerous than translational ones, so long as the main part of the gravitational force is sequestered in the curvature -- we discuss this idea further in the context of some new computer algebra tools in~\cite{crunchy}.
Rather than attempting to constrain $\PiP[\ovl{i}]{A1m}$, we might think to target the conjugate field. By inspecting~\eqref{simspione} and~\eqref{courel}, it would seem that $\cbet{2}\neq 0$ is a possible option. Once again we expect this to impact the $1^-$ sector, since $\bet{2}=0$, while collateral damage seems to be confined to the $0^+$ translational sector. 
On closer inspection, however, we notice from~\eqref{ghoqor8} that the condition $\cbet{2}\neq 0$ is not actually deactivating the bad sector: it is relating velocities and gradients within the torsion.
Despite this, we find it useful to proceed with the analysis.
The `collateral' $0^+$ sector is non-conjugate, in that the $0^+$ spin state is represented only in the velocity-dependent or nonphysical components of the torsion, and not in the canonical part $\tensor{\mathcal{T}}{^i_{\ovl{kl}}}$. For this reason, there will be no parallel secondary to worry about, which makes for a less complex analysis. Moreover, and regardless of the unitarity of the model, we believe the $\cbet{2}\neq 0$ configuration to amply demonstrate how multipliers might be used to modify the linear-to-nonlinear transition.

The PiCs of the new theory are 
\begin{subequations}
  \begin{align}
    \pic[]{B0p}&\equiv \frac{1}{J}\PiP[]{B0p}+2\planck^2\cbet{2}\ncTLambda[]{B0p}\approx 0,
    \label{pic1}
    \\
    \pic[\ovl{i}]{B1m}&\equiv \frac{1}{J}\PiP[\ovl{i}]{B1m}+\frac{2}{3}\planck^2\cbet{2}\left(\cTLambda[\ovl{i}]{B1m}+\ncTLambda[\ovl{i}]{B1m}\right)\approx 0,\label{pic2}
    \\
    \pic[\ovl{ij}]{B2p}&\equiv \frac{1}{J}\PiP[\ovl{ij}]{B2p}\approx 0,
    \\
    \pic[]{A0p}&\equiv \frac{1}{J}\PiP[]{A0p}+3\planck^2\alp{0}\approx 0,
    \\
    \pic[]{A0m}&\equiv \frac{1}{J}\PiP[]{A0m}\approx 0,
    \\
    \pic[\ovl{ij}]{A2p}&\equiv \frac{1}{J}\PiP[\ovl{ij}]{A2p}\approx 0,
    \\
    \pic[\ovl{ijk}]{A2m}&\equiv \frac{1}{J}\PiP[\ovl{ijk}]{A2m}\approx 0,
    \label{picm1}
  \end{align}
\end{subequations}
where~\cref{pic1,pic2} go over to their original counterparts in the theory without multipliers by taking $\cbet{2}\to 0$,
and there is an extra pair of primaries stemming from~\eqref{newp}
  \begin{equation}
	  \cTpic[\ovl{i}]{B1m}\equiv \cTPiP[\ovl{i}]{B1m}+\ncTPiP[]{B0p}\approx
    \ncTpic[]{B0p}\equiv \ncTPiP[]{B0p}\approx 0.
    \label{pvb}
  \end{equation}
  Note that we are using the `variable-index' notation in~\cref{ireppl}.

  Conveniently, we note that the nonlinear Poisson brackets between the translational and rotational (i.e. non-geometric) primaries are the same even in the $\cbet{2}\to 0$ limit. We provide these commutators in~\crefrange{B0pA0p}{B2pA2m}.

\subsection{The new super-Hamiltonian}

For the purpose of evaluating velocities, it is important to understand the new super-Hamiltonian described in~\eqref{ghoul}. By imposing the conditions~\eqref{simspione} and restricting to the PiC shell in~\crefrange{pic1}{picm1} among the quadratic terms (which can always be done by redefining the Hamiltonian multipliers), we obtain
  \begin{align}
    \mathcal{H}_{\perp}&\approx
    \frac{\etau{\ovl{ij}}\PiP[\ovl{i}]{A1m}\PiP[\ovl{j}]{A1m}}{16\alp{5}J}
    +\frac{\etau{\ovl{ik}}\etau{\ovl{jl}}\PiP[\ovl{ij}]{A1p}\PiP[\ovl{kl}]{A1p}}{8\alp{5}J}
    +\frac{3\etau{\ovl{ik}}\etau{\ovl{jl}}\PiP[\ovl{ij}]{B1p}\PiP[\ovl{kl}]{B1p}}{16\bet{3}J}
    \nonumber\\
    &\ \ 
    +\frac{\alp{0}\planck^2J\cR[]{A0p}}{2}
    -\frac{\etau{\ovl{ij}}\PiP[\ovl{i}]{A1m}\cR[\ovl{j}]{A1m}}{2}
    +\etau{\ovl{ik}}\etau{\ovl{jl}}\PiP[\ovl{ij}]{A1p}\cR[\ovl{kl}]{A1p}
    \nonumber\\
    &\ \ 
    +\frac{\bet{3}\planck^2J\cT[]{A0m}^2}{6}
    -\frac{2\cbet{2}\planck^2j\etau{\ovl{ij}}\cTLambda[\ovl{i}]{B1m}\cT[\ovl{j}]{B1m}}{3}
    \nonumber\\
    &\ \  
    -\frac{2\cbet{2}\planck^2j\etau{\ovl{ij}}\ncTLambda[\ovl{i}]{B1m}\cT[\ovl{j}]{B1m}}{3}
    +\frac{\etau{\ovl{ik}}\etau{\ovl{jl}}\PiP[\ovl{ij}]{B1p}\cT[\ovl{kl}]{B1p}}{2}
    \nonumber\\
    &\ \  
    +\frac{16\bet{3}\planck^2 J\etau{\ovl{ik}}\etau{\ovl{jl}}\etau{\ovl{mn}}\cT[\ovl{ijm}]{A2m}\cT[\ovl{kln}]{A2m}}{27}
    \nonumber\\
    &\ \  
-\tensor{n}{^k}\tensor{D}{_\alpha}\tensor{\pi}{_k^\alpha}
\approx 0.
\label{grandmaster}
  \end{align}

  We see in~\eqref{grandmaster} the origin of the ghost nature of the strongly coupled $1^-$ sector. If the sign of $\alp{5}$ is fixed to cause the $1^+$ momentum $\PiP[\ovl{ij}]{A1p}$ to enter with positive energy, then the same cannot apply to the quadratic term built from the parity-odd $1^-$ momentum $\PiP[\ovl{i}]{A1m}$. Hence, the strong coupling of the $1^-$ mode introduces a nonlinear ghost. We now proceed to the Dirac--Bergmann algorithm.

\subsection{Consistency of geometric primaries}

In what follows we will use $\arb{}(x,y,\ldots |u,v,\ldots)$ to indicate a function linear in $x$, $y$ and its other (arbitrarily indexed) arguments, with coefficients depending on $\foli{i}$, $\etad{\ovl{ij}}$, $\epsd{\ovl{ijk}}$ and $J$ -- i.e. quantities persisting in the linearised theory -- along with $\planck$ and any of the couplings in~\eqref{neocon}. The arguments $u$, $v$, etc., will be nonlinear corrections to those coefficients (if any). Note that wherever we use $\arb{}$, we are also asserting implicitly that these coefficients may be straightforwardly determined, though the calculation may be lengthy.

Beginning with the $0^+$ sector, the consistency of $\ncTpic[]{B0p}$ concerns only the bracket
\begin{equation}
  \left\{
\ncTpic[]{B0p}
,
\pic[]{B0p}
  \right\}\approx
  -6\cbet{2}\planck^2
  \delta^3,
\end{equation}
and is satisfied when we fix the multiplier
\begin{equation}
  \mul[]{B0p}\approx
  0,
  \label{cabbob}
\end{equation}
irrespective of whether we consider the linearised or nonlinear theories.

In the $1^-$ sector, the consistency of $\cTpic[\ovl{i}]{B1m}$ involves the bracket
\begin{equation}
  \left\{
\cTpic[\ovl{i}]{B1m}
,
\pic[\ovl{l}]{B1m}
  \right\}\approx
  \arb{}\big(
   1 
  \ \big|\cdot
\big)
  \delta^3,
  \label{cabbab}
\end{equation}
and is satisfied when we fix, with reference to~\eqref{mulnew}, the multiplier
\begin{equation}
  \mul[\ovl{i}]{B1m}
  \approx
  \arb{}\big(
  \cT[\ovl{i}]{B1m}
  |\cdot
\big).
\label{cabbib}
\end{equation}
All the geometric primaries are then consistent with the determined multipliers in~\cref{cabbob,cabbib}.

\subsection{Consistency of rotational primaries}

Having dealt with the geometric primaries, we now return to the part of the analysis familiar from~\cite{2002IJMPD..11..747Y} and~\cref{simpleA1p}: the consistency of the primaries of gravitational gauge fields in~\crefrange{pic1}{picm1}. We begin again with the conjugate part of the rotational sector. The consistency of the $0^+$ part $\pic[]{A0p}$ involves the brackets~\eqref{B0pA0p} and~\eqref{B1mA0p}. Noting the conjugate commutator~\eqref{B0pA0p}, we remember that this was previously used to determine $\mull[]{B0p}$ in the linearised theory\footnote{Recall from~\cite{2002IJMPD..11..747Y,chapter4,mythesis} that the symbol ($\flat$) is used to denote linearisation.}. This time around, the geometric primaries have gotten there first (by our choice of ordering). Despite the conjugacy of the $0^+$ PiC, we are thus forced to admit a secondary if-constraint (SiC)
\begin{equation}
  \sic[]{A0p}\equiv \arb{}\big(
  \textstyle\int\mathrm{d}^3x N\big\{
    \pic[]{A0p}
  ,
    \mathcal{H}_{\perp}
  \big\}
    ,\
    \mul[\ovl{i}]{B1m}
    \cdot
    \PiP[\ovl{j}]{A1m}
  \ \big|\cdot
\big)\approx 0.
\label{cibibi}
\end{equation}
The notation ($\cdot$) in the first set of arguments in~\eqref{cibibi} denotes one or more suitably indexed and symmetrized products (in the second set of arguments, it indicates that there are no corrections to consider).
Without immediately evaluating this, we move on to the conjugate $2^+$ sector. This is a somewhat more familiar setup, in which the only brackets are~\cref{B1mA2p,B2pA2p}.
We may then use this setup to solve for the conjugate multiplier
\begin{equation}
  \mul[\ovl{ij}]{B2p}\approx \arb{}\big(
  \textstyle\int\mathrm{d}^3x N\big\{
    \pic[\ovl{ij}]{A2p}
  ,\
    \mathcal{H}_{\perp}
  \big\}
    ,\
    \mul[\ovl{i}]{B1m}
    \cdot
    \PiP[\ovl{j}]{A1m}
  \ \big| \
  \PiP[\ovl{ij}]{A1p}
\big),
\label{cobbeb}
\end{equation}
thereby ensuring the consistency of $\pic[\ovl{ij}]{A2p}$, regardless of whether the theory is linearised or not.

The parity-odd rotational sectors are not conjugate. In~\eqref{B1mA0m} we encounter yet another strictly nonlinear commutator with the translational $1^-$ sector, and since $\mul[\ovl{i}]{B1m}$ was solved in~\eqref{cabbib} we must emulate the technique of the linearised theory and again construct a SIC 
\begin{equation}
  \sic[]{A0m}\equiv 
  \arb{}\big(
  \textstyle\int\mathrm{d}^3x N\big\{
    \pic[]{A0m}
  ,
    \mathcal{H}_{\perp}
  \big\}
    ,\
    \mul[\ovl{j}]{B1m}\cdot\PiP[\ovl{pq}]{A1p}
  \ \big|\cdot
\big)
\approx 0,
\label{cebebe}
\end{equation}
whose evaluation we again defer. The same situation applies for the $2^-$ sector: from~\cref{B1mA2m,B2pA2m}, and noticing that the geometric multipliers already allowed us to solve $\mul[\ovl{ij}]{B2p}$ in~\eqref{cobbeb}, we must construct
\begin{equation}
  \begin{aligned}
    \sic[\ovl{ijk}]{A2m}\equiv 
  \arb{}\big(&
  \textstyle\int\mathrm{d}^3x N\big\{
    \pic[\ovl{ijk}]{A2m}
  ,
    \mathcal{H}_{\perp}
  \big\}
    ,
    \\
    &\ \ \ 
    \mul[{\ovl{i}}]{B1m}\cdot \PiP[\ovl{jk}]{A1p}
    ,\
    \mul[{\ovl{ij}}]{B2p}\cdot \PiP[\ovl{k}]{A1m}
  \ \big|\cdot
\big)
\approx 0.
  \end{aligned}
  \label{cubabo}
\end{equation}
So far we have fixed the consistencies of all the usual rotational primaries in ways which do not qualitatively change as we move from the linear to nonlinear theories. We have tried to use mostly the same technique as for the linearised theory without geometric multipliers: that of playing off sectors with the same $J^P$ against each other.

\subsection{Consistency of rotational secondaries}

We must not, however, forget the secondaries in~\cref{cibibi,cebebe,cubabo}, which are not yet consistent. In the case of the $0^-$ and $2^-$ sectors, we can tentatively assume that the situation is the same as for the linearised theory \emph{without} multipliers: the natural conjugates of these secondaries will be their own primaries, allowing us to obtain
\begin{subequations}
  \begin{align}
  \mul[]{A0m}&\approx \arb{}\big(
  \textstyle\int\mathrm{d}^3x N\big\{
    \sic[]{A0m}
  ,\
    \mathcal{H}_{\perp}
  \big\}
  ,\
  \ldots
  |
  \ldots
\big),
\label{covel}
\\
  \mul[\ovl{ijk}]{A2m}&\approx \arb{}\big(
  \textstyle\int\mathrm{d}^3x N\big\{
    \sic[\ovl{ijk}]{A2m}
  ,\
    \mathcal{H}_{\perp}
  \big\}
  ,\
  \ldots
  |
  \ldots
\big),
\label{coval}
  \end{align}
\end{subequations}
where we allow some space to parameterise our ignorance of any other commutators which may arise. Thus, while we expect to be able to construct~\cref{covel,coval} in the full nonlinear theory, we do not expect them to explicitly determine $\mul[]{A0m}$ and $\mul[\ovl{ijk}]{A2m}$, due to emergent dependencies on yet-undetermined multipliers.

What about the novel $0^+$ secondary? The linearised theory without multipliers does not suggest to us the commutators of $\sic[]{A0p}$, so to discover these we must obtain an explicit formula from~\eqref{cibibi}. We defer the full expression to~\cref{prospekt}, but for the moment we notice that only the $\textstyle\int\mathrm{d}^3x N\big\{\pic[]{A0p},\mathcal{H}_{\perp}\big\}$ term will contribute in the linearised theory to $\sicl[]{A0p}$. Specifically, if we focus on the non-quadratic part in~\eqref{ghoul}, and refer to the PiC shell condition in~\eqref{pic1} we have
  \begin{align}
  \left\{
    \pic[]{A0p}
,
-\tensor{n}{^k}\tensor{D}{_\alpha}\tensor{\pi}{_k^\alpha}
  \right\}&\approx
  \Big[
    2\alp{0}\planck^2\tensor{\mathcal{D}}{_{\ovl{l}}}\foliu{l}
    +\frac{1}{2J}\etau{\ovl{ij}}\etau{\ovl{kl}}\PiP[\ovl{ik}]{A1p}\cT[\ovl{jl}]{B1p}
    \nonumber\\
    &
    -2\cbet{2}\planck^2\ncTLambda[]{B0p}
  \Big]
  \delta^3
  +\delta^3\etau{\ovl{ij}}\PiP[\ovl{i}]{A1m}\tensor{\mathcal{D}}{_{\ovl{j}}}
  ,
  \label{strangemass}
  \end{align}
  where we use the Dirac function gradient in~\cref{surfic}. On this basis it would then appear that $\big\{\sicl[]{A0p},\ncTpicl[]{B0p}\big\}\not\approx 0$. This should also apply in the nonlinear theory, so we can follow~\cref{covel,coval} in writing
  \begin{align}
  \ncTmul[]{B0p}&\approx \arb{}\big(
  \textstyle\int\mathrm{d}^3x N\big\{
    \sic[]{A0p}
  ,\
    \mathcal{H}_{\perp}
  \big\}
  ,\
  \ldots
  |
  \ldots
\big).
\label{covol}
  \end{align}
  At least for the linearised case, it would seem that the consistencies of all the secondaries raised by the rotational sector are exactly absorbed by previously undetermined Hamiltonain multipliers: the algorithm in the this sector is then terminated.

\subsection{Consistency of translational primaries}

  Moving over to the translational primaries, we find that we can absorb all their consistency conditions exactly by determining the conjugate rotational or translational-geometric Hamiltonian multipliers. Referring first to~\cref{B0pA0p,cabbob}, then to \cref{B2pB2p,B2pA2p,B2pA2m}, we write
\begin{subequations}
  \begin{align}
  \mul[]{A0p}&\approx \arb{}\big(
  \textstyle\int\mathrm{d}^3x N\big\{
    \pic[]{B0p}
  ,\
    \mathcal{H}_{\perp}
  \big\}
  ,\
  \ncTmul[]{B0p}
  \ \big|\cdot
\big),
\label{caval}
\\
  \mul[\ovl{ij}]{A2p}&\approx \arb{}\big(
  \textstyle\int\mathrm{d}^3x N\big\{
    \pic[\ovl{ij}]{B2p}
  ,\
    \mathcal{H}_{\perp}
  \big\}
  ,
  \nonumber\\
  &
  \quad
  \quad
  \quad
  \quad
  \mul[\ovl{ij}]{B2p}\cdot\PiP[\ovl{kl}]{B1p}
  ,\
  \mul[\ovl{ijk}]{A2m}\cdot\PiP[\ovl{l}]{A1m}
  \ \big| \
  \PiP[\ovl{ij}]{A1p}
\big).
\label{civil}
  \end{align}
\end{subequations}
As before, these solutions are still available as we extend to the nonlinear theory, however they do rely on the continued success of~\cref{coval,covol} for an explicit solution.
Finally we tackle the problematic $1^-$ sector, the only constraint whose consistency is not yet established. Noticing that~\cref{B1mB1m,B1mA0p,B1mA0m,B1mA2p,B1mA2m} are all nonlinear commutators, it seems advantageous that the only linear commutator~\eqref{cabbab} allows us to solve for the remaining Hamiltonian multiplier
  \begin{align}
  \cTmul[\ovl{i}]{B1m}&\approx \arb{}\big(
  \textstyle\int\mathrm{d}^3x N\big\{
    \pic[\ovl{i}]{B1m}
  ,\
    \mathcal{H}_{\perp}
  \big\}
  ,\ 
  \mul[\ovl{i}]{B1m}\cdot\PiP[\ovl{jk}]{B1p}
  ,\
  \mul[]{A0p}\cdot\PiP[\ovl{j}]{A1m}
  ,
  \nonumber\\
  &
  \quad
  \quad
  \quad
  \quad
  \mul[]{A0m}\cdot\PiP[\ovl{ij}]{A1p}
  ,\
  \mul[\ovl{ij}]{A2p}\cdot\PiP[\ovl{k}]{A1m}
  ,\
  \mul[\ovl{ijk}]{A2m}\cdot\PiP[\ovl{lm}]{A1p}
  \ \big|\cdot
\big).
\label{cevel}
  \end{align}
  This, too, is expected to hold for the nonlinear theory, but it does rely on explicit solutions from the system~\cref{covel,coval,covol}. All the translational consistencies are absorbed by Hamiltonian multipliers. No constraints remain, and the Dirac--Bergmann algorithm is terminated.

  \subsection{Nonlinear prospects}\label{prospekt}

  In summary, we have obtained all Hamiltonian multipliers. All the primaries are SC, and we propose to construct three sets of SC secondaries in $\sic[]{A0p}$, $\sic[]{A0m}$ and $\sic[\ovl{ijk}]{A2m}$. 

  Wherever possible in the calculations above, we have ensured that Hamiltonian multipliers do not need to be solved simultaneously, the RHS of~\cref{cabbob,cabbib,cobbeb,caval,civil,cevel} each depending in turn on predetermined quantities. This convenient pattern could be broken within~\cref{covel,coval,covol}, which are not explicitly obtained, and which we tentatively assume (referring back to the traditional case~\cite{2002IJMPD..11..747Y}) not to be a singular system in the Hamiltonian multipliers. 

  Accordingly, some changes to the order and structure of these solutions are expected as we pass from the linearised to the nonlinear theory. To see this, we use~\eqref{grandmaster} to determine the \emph{nonlinear} $0^+$ SiC as follows
    \begin{align}
    \sic[]{A0p}&\approx
    -2\cbet{2}\planck^2 \ncTLambda[]{B0p}
    -\etau{\ovl{ij}}\tensor{\mathcal{D}}{_{\ovl{i}}}\bigg(\frac{\PiP[\ovl{j}]{A1m}}{J}\bigg)
    -\frac{\foliu{l}\etau{\ovl{ij}}\tensor{\mathcal{D}}{_{\ovl{i}}}\PiP[\ovl{jl}]{A1p}}{J}
    \nonumber\\
    &\ \ 
    -\frac{\etau{\ovl{ik}}\PiP[\ovl{i}]{A1m}\cT[\ovl{k}]{B1m}}{2J}
    +\frac{\etau{\ovl{ik}}\etau{\ovl{jl}}\PiP[\ovl{ij}]{A1p}\cT[\ovl{kl}]{B1p}}{J}
    +\frac{3\etau{\ovl{ik}}\etau{\ovl{jl}}\PiP[\ovl{ij}]{A1p}\PiP[\ovl{kl}]{B1p}}{8\bet{3}\planck^2J}
    \nonumber\\
    &\ \ 
    +\arb{}\big(
      \mul[]{B0p}
      ,\
      \mul[\ovl{i}]{B1m}\cdot\PiP[\ovl{j}]{A1m}
      \ \big|\cdot
    \big)
    \approx 0,
  \end{align}
  where we do not need to obtain the precise coupling of the two Hamiltonian multipliers which are tangled up in the quantity. We see that the guess in~\eqref{strangemass} is borne out, so that $\big\{\sicl[]{A0p},\ncTpicl[]{B0p}\big\}\not\approx 0$ can be obtained from~\eqref{cabbob}. However this is not the only commutator, as we readily find for the nonlinear case
  \begin{equation}
    \lim_{\mul[]{B0p},\ \mul[\ovl{i}]{B1m}\to 0}
    \left\{
      \pic[\ovl{ij}]{A2p}
      ,
      \sic[]{A0p}
    \right\}\approx
    \frac{3\etau{\ovl{kl}}\PiP[\langle\ovl{i}|\ovl{k}]{A1p}\PiP[\ovl{l}|\ovl{j}\rangle]{A1p}}{8\bet{3}\planck^2J^3}.
    \label{blight}
  \end{equation}
  The result in~\eqref{blight} suggests that~\cref{covol,civil} will become co-dependent, scrambling the order of the linear solution method. The general system is expected to be more complicated.

\begin{figure}[t!]
  \center
  \includegraphics[width=\linewidth]{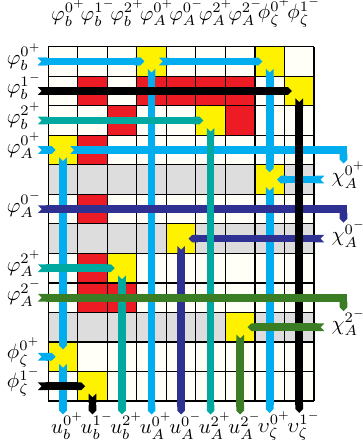}
	\caption{\label{ConstraintAlgebra} 
	The constraint algebra of the theory defined by~\eqref{simspione} with $\cbet{2}\neq 0$, as it appears in its simplest form on the constraint shell. 
	All Hamiltonian multipliers $\tensor*{u}{^{J^P}_{b}}$, $\tensor*{u}{^{J^P}_{A}}$ and $\tensor*{\upsilon}{^{J^P}_{\zeta}}$ are eventually determined in~\cref{cabbob,cabbib,cobbeb,covel,coval,covol,caval,civil,cevel} by satisfying the consistency conditions of the primary constraints $\tensor*{\varphi}{^{J^P}_{b}}$ and $\tensor*{\varphi}{^{J^P}_{A}}$ in~\crefrange{pic1}{picm1} and $\tensor*{\phi}{^{J^P}_{\zeta}}$ in~\eqref{pvb} (lines coloured by $J^P$) via abundant order-unity Poisson brackets (yellow squares). For ease of notation in this figure alone, we label the $J^P$ explicitly and drop the combination of Lorentz indices and accents used in the body of text. Perturbative brackets (red squares) do not reduce the number of induced constraints $\tensor*{\chi}{^{J^P}_{A}}$, which would otherwise indicate strong coupling. Some brackets were not computed for this work (gray squares). Subtracting the $\sum_{J}2J+1$ constrained multiplicities leaves six extra degrees of freedom.
	}
\end{figure}

  In order to calculate the d.o.f, we work with our picture of the linearised theory and assume (by the example of the original PGT in~\eqref{pgtqp}), that the sSFCs in~\eqref{suresecondaries} always remain in the final reckoning both FC and independent as geometric multipliers are introduced. This postulate could be tested by investigating the constraint algebra~\cite{blagojevic2002gravitation}. In that case we obtain\footnote{Using basically the same labelling scheme in~\cite{chapter4,mythesis,2002IJMPD..11..747Y}, where if-constraints iP(S)F(S)C are primary(secondary) first(second) class.}
\begin{equation}
\begin{aligned}
  8&=\smash{\frac{1}{2}}\big(80+2\times(3+1)[\cbet{2}-\text{multipliers}]
    \\
    &\ \ \ \ \ \ \ 
    -2\times 10[\text{sPFC}]-2\times 10[\text{sSFC}]
    \\
    &\ \ \ \ \ \ \ 
    -(1+3+5+1+1+5+5+3+1)[\text{iPSC}]
    \\
    &\ \ \ \ \ \ \ 
    -(1+1+5)[\text{iSSC}]
  \big).
    \label{twoeights}
\end{aligned}
\end{equation}
It would seem that \emph{six} d.o.f are now moving independently of the graviton: this structure is shown in~\cref{ConstraintAlgebra}.

In case of a breakdown of the linear solution, as warned about above, we use the very general method of Yo and Nester~\cite{2002IJMPD..11..747Y} when checking the nonlinear d.o.f. We also assume the properties of the sSFCs to be preserved. The evaluation of $\mul[]{B0p}$ and $\mul[\ovl{i}]{B1m}$ ought to remain safe at all orders. Generically, the consistencies of the rotational primaries $\pic[]{A0p}$, $\pic[]{A0m}$, $\pic[\ovl{ij}]{A2p}$ and $\pic[\ovl{ijk}]{A2m}$ should determine $\mul[\ovl{ij}]{B2p}$ and `overflow' into seven SiCs $\chi_{[7]}$. The consistencies of $\pic[\ovl{ij}]{B2p}$ and $\chi_{[7]}$ together with $\pic[]{B0p}$ and $\pic[\ovl{i}]{B1m}$ collectively determine $\mul[]{A0p}$, $\mul[]{A0m}$, $\mul[\ovl{ij}]{A2p}$, $\mul[\ovl{ijk}]{A2m}$ and also $\cTmul[\ovl{i}]{B1m}$ and $\ncTmul[]{B0p}$. The situation may involve cross-talk between $J^P$ sectors, but it seems to retain the eight d.o.f in~\eqref{twoeights} as we move away from the torsion-free, Minkowski background.

\section{closing remarks}\label{chapter5conclusions}

In this paper we have extended the original Poincar\'e gauge theory (PGT) in~\eqref{pgtqp} by introducing in~\eqref{neocon} a complete set of \emph{geometric multiplier} fields, which deactivate various parts of the Riemann--Cartan curvature and the torsion.
Teleparallel gravity is a special case of this extension.
A chronic problem with the original PGT is the appearance of nonlinear commutators between primary constraints: these cause a departure from the linearised constraint structure which is manifest as a strong coupling problem.
This problem is especially severe in purely quadratic theories (which can exhibit otherwise viable phenomenology without any Einstein--Hilbert term~\cite{chapter2,chapter3,chapter4,mythesis}) since the removal of the Einstein--Hilbert term leads to sparse commutators at linear order. 

We have examined the effects of geometric multipliers on the canonical analysis. 
New primary and secondary constraints are produced, which typically fail to commute with the original primaries at linear order.
Heuristically, the efficacy of this mechanism will be connected to the fact that the new fields in~\eqref{neocon} are multipliers: propagating d.o.fs would tend to increase with any new kinetic terms, offsetting any benefits.

We experimented with this approach on a non-viable example of the PGT, in which the Einstein--Cartan term is joined by quadratic torsion and curvature invariants so as to introduce a unitary, massive spin-parity ($J^P$) $1^+$ torsion vector in the linearised particle spectrum. That theory was known to contain a strongly coupled $1^-$ ghost in its nonlinear completion. We imposed a minimal geometric multiplier, which affects the bad sector so as to equate the velocities and gradients within the $1^-$ part of the torsion tensor. 

The modified particle spectrum in our example is not likely to be healthy before or after linearisation. Rather than a transition from five to eight d.o.f, we seem always to have eight. Even if the assumptions made during the evaluation are justified, we provide no reason to be optimistic about the resulting unitarity. 
It is not surprising that it is hard to use geometric multipliers to repair pre-existing theories. 
By so changing the linear constraint structure, unwanted modes may be activated and the unitarity destroyed.
Such also was the case in~\cite{mythesis} for the purely quadratic theory developed in~\cite{chapter2,chapter3}: those same multipliers which do not interfere with the viable cosmology or gravitational waves of the original theory were shown to induce classical ghosts on the flat, torsion-free background. In that case the situation was not so critical, since the relevant cosmological background of that theory is thought instead to contain a constant, finite torsion condensate.

Notwithstanding the possible ghost and tachyon content of our example particle spectrum, the analysis shows that multiplier configurations may offer a novel route to softening the nonlinear transition in non-Riemannian theories, and so combat strong coupling. Solving for unitarity without strong coupling represents a major undertaking~\cite{Lin1,Lin2,Lin3}. In the companion paper~\cite{crunchy} we present computer algebra tools for performing the requisite canonical analysis at scale.

\begin{acknowledgements}
  This work was performed using resources provided by the Cambridge Service for Data Driven Discovery (CSD3) operated by the University of Cambridge Research Computing Service (\href{www.csd3.cam.ac.uk}{www.csd3.cam.ac.uk}), provided by Dell EMC and Intel using Tier-2 funding from the Engineering and Physical Sciences Research Council (capital grant EP/T022159/1), and DiRAC funding from the Science and Technology Facilities Council (\href{www.dirac.ac.uk}{www.dirac.ac.uk}).

  This manuscript was improved by the kind suggestions of Amel Durakovi\'c.

  I would like to thank Milutin Blagojevi\'c for his cogent and helpful correspondence in October of 2021, and Tom Z\l o\'snik for useful conversations.

I am grateful for the kind hospitality of Leiden University and the Lorentz Institute, and the support of Girton College, Cambridge.

\end{acknowledgements}

\bibliographystyle{apsrev4-1}
\bibliography{bibliography}
\appendix

\section{Irreducible decomposition}\label{ireppl}

In this appendix, we detail the notation used in~\cite{chapter4,mythesis,crunchy} for the $\sothree$ irreps of various quantities.
The translational parallel momentum decomposes as
\begin{subequations}
\begin{align}
  \tensor{\hat{\pi}}{_{k\ovl{l}}}&=\tensor{\hat\pi}{_{\ovl{kl}}}+\tensor{n}{_k}\PiP[\ovl{l}]{B1m},\label{traper}\\
  \tensor{\hat{\pi}}{_{\ovl{kl}}}&=\frac{1}{3}\etad{\ovl{kl}}\PiP{B0p}+\PiP[\ovl{kl}]{B1p}+\PiP[\ovl{kl}]{B2p},
  \label{trapar}
\end{align}
\end{subequations}
where in~\eqref{traper} there is a $1^-$ irrep, and a term which further decomposes in~\eqref{trapar} to give $0^+$, antisymmetric $1^+$, and symmetric (but traceless) $2^+$ irreps. 
Similarly, the rotational parallel momentum is
\begin{subequations}
\begin{align}
  \tensor{\hat{\pi}}{_{kl\ovl{m}}}&=\tensor{\hat{\pi}}{_{\ovl{klm}}}+2\tensor{n}{_{[k}}\tensor{\hat{\pi}}{_{\perp\ovl{l}]\ovl{m}}},\\
  \tensor{\hat{\pi}}{_{\perp\ovl{kl}}}&=\frac{1}{3}\etad{\ovl{kl}}\PiP{A0p}+\PiP[\ovl{kl}]{A1p}+\PiP[\ovl{kl}]{A2p},\label{rotper}\\
  \tensor{\hat{\pi}}{_{\ovl{klm}}}&=\frac{1}{6}\epsd{\ovl{klm}}\PiP{A0m}+\PiP[[\ovl{k}]{A1m}\etad{{{\ovl{l}]\ovl{m}}}}+\frac{4}{3}\PiP[\ovl{klm}]{A2m},
  \label{rotpar}
\end{align}
\end{subequations}
with $0^+$, $1^+$ and $2^+$ modes in~\eqref{rotper}, and $0^-$, $1^-$ and $2^-$ modes in~\eqref{rotpar}.

The field strengths are decomposed into parallel and perpendicular parts
\begin{subequations}
\begin{align}
  \tensor{\mathcal{R}}{_{ijkl}}&=\tensor{\mathcal{R}}{_{ij\ovl{kl}}}+2\tensor{n}{_{[k|}}\tensor{\mathcal{R}}{_{ij\perp|\ovl{l}]}},\label{koffl}
 \\
 \tensor{\mathcal{T}}{_{ikl}}&=\tensor{\mathcal{T}}{_{i\ovl{kl}}}+2\tensor{n}{_{[k|}}\tensor{\mathcal{T}}{_{i\perp|\ovl{l}]}}.\label{foffl}
\end{align}
\end{subequations}
The parallel $\tensor{\mathcal{R}}{_{ij\ovl{kl}}}$ contains the $0^+$ part $\cR[]{A0p}$, $1^+$ part $\cR[\ovl{ij}]{A1p}$, $2^+$ part $\cR[\ovl{ij}]{A2p}$, $0^-$ part $\cR[]{A0m}$, $1^-$ part $\cR[\ovl{i}]{A1m}$ and $2^-$ part $\cR[\ovl{ijk}]{A2m}$. The ($\langle\cdot \rangle$) notation indicates the symmetric-traceless operation. 
The perpendicular $\tensor{\mathcal{R}}{_{ij\perp\ovl{l}}}$ contains the $0^+$ part $\ncR[]{A0p}$, $1^+$ part $\ncR[\ovl{ij}]{A1p}$, $2^+$ part $\ncR[\ovl{ij}]{A2p}$, $0^-$ part $\ncR[]{A0m}$, $1^-$ part $\ncR[\ovl{i}]{A1m}$ and $2^-$ part $\ncR[\ovl{ijk}]{A2m}$.

The parallel $\tensor{\mathcal{T}}{_{i\ovl{kl}}}$ contains the $0^-$ part $\cT[]{A0m}$, $1^+$ part $\cT[\ovl{ij}]{B1p}$, $1^-$ part $\cT[\ovl{i}]{B1m}$ and $2^-$ part $\cT[\ovl{ijk}]{A2m}$. The perpendicular $\tensor{\mathcal{T}}{_{i\perp\ovl{l}}}$ contains the $0^+$ part $\ncT[]{B0p}$, $1^+$ part $\ncT[\ovl{ij}]{B1p}$, $1^-$ part $\ncT[\ovl{i}]{B1m}$ and $2^+$ part $\ncT[\ovl{ij}]{B2p}$.

Our conventions for the $\sothree$ irreps of other quantities whose Lorentz indices have the same structure, are to recycle the various expressions above, replacing only the symbols $\hat{\pi}$, $\mathcal{R}$, $\mathcal{T}$, etc.

\section{The transfer couplings}\label{mulcou}
We provide in this appendix translations of the transfer couplings from~\cref{manymany} into the formalisms set out in~\cite{chapter4}.
The first half of the transfer couplings in~\eqref{manymany} are found to be
\begin{equation}
  \begin{aligned}
    \calpm[0^+]{\parallel\parallel}&\equiv\frac{1}{2}(\calp{4}+\calp{6}), & \calpm[0^-]{\parallel\parallel}&\equiv\frac{1}{2}(\calp{2}+\calp{3}), \\
    \calpm[1^+]{\parallel\parallel}&\equiv-\frac{1}{2}(\calp{2}+\calp{5}), & \calpm[1^-]{\parallel\parallel}&\equiv\frac{1}{2}(\calp{4}+\calp{5}), \\
    \calpm[2^+]{\parallel\parallel}&\equiv\frac{1}{2}(\calp{1}+\calp{4}), & \calpm[2^-]{\parallel\parallel}&\equiv\frac{1}{2}(\calp{1}+\calp{2}), \\
    \calpm[0^+]{\perp\parallel}&\equiv-\frac{1}{4}(\calp{4}-\calp{6}), & \calpm[0^-]{\perp\parallel}&\equiv\frac{1}{2}(\calp{2}-\calp{3}), \\
    \calpm[1^+]{\perp\parallel}&\equiv-\frac{1}{2}(\calp{2}-\calp{5}), & \calpm[1^-]{\perp\parallel}&\equiv\frac{1}{2}(\calp{4}-\calp{5}), \\
    \calpm[2^+]{\perp\parallel}&\equiv\frac{1}{2}(\calp{1}-\calp{4}), & \calpm[2^-]{\perp\parallel}&\equiv-\frac{1}{2}(\calp{1}-\calp{2}),
  \end{aligned}
  \label{courel}
\end{equation}
and the remaining couplings are \emph{mostly} found using the rules $\calpm[A]{\perp\perp}\equiv\frac{1}{2}\calpm[A]{\parallel\parallel}$ and $\calpm[A]{\parallel\perp}\equiv\frac{1}{2}\calpm[A]{\perp\parallel}$, with the three exceptions $\calpm[1^+]{\perp\perp}\equiv-\frac{1}{2}\calpm[1^+]{\parallel\parallel}$, $\calpm[1^+]{\parallel\perp}\equiv-\frac{1}{2}\calpm[1^+]{\perp\parallel}$ and $\calpm[0^+]{\perp\parallel}\equiv \frac{1}{2}\calpm[0^+]{\parallel\perp}$, and these quirks just result from the `human' normalisation of the $\othree$ representations.
It goes without saying that a precisely equivalent formulation can be constructed for the couplings $\{\alp{I}\}$.
The translational transfer couplings are
\begin{equation}
  \begin{aligned}
    \cbetm[0^+]{\parallel\parallel}&\equiv 0, & \cbetm[0^-]{\parallel\parallel}&\equiv\frac{1}{6}\cbet{3}, \\
    \cbetm[1^+]{\parallel\parallel}&\equiv\frac{1}{3}(2\cbet{1}+\cbet{3}), & \cbetm[1^-]{\parallel\parallel}&\equiv\frac{1}{3}(\cbet{1}+2\cbet{2}), \\
    \cbetm[2^+]{\parallel\parallel}&\equiv0, & \cbetm[2^-]{\parallel\parallel}&\equiv\cbet{1}, \\
    \cbetm[0^+]{\perp\parallel}&\equiv0, & \cbetm[0^-]{\perp\parallel}&\equiv0, \\
    \cbetm[1^+]{\perp\parallel}&\equiv-\frac{1}{3}(\cbet{1}-\cbet{3}), & \cbetm[1^-]{\perp\parallel}&\equiv-\frac{1}{3}(\cbet{1}-\cbet{2}), \\
    \cbetm[2^+]{\perp\parallel}&\equiv0, & \cbetm[2^-]{\perp\parallel}&\equiv0, \\
    \cbetm[0^+]{\perp\perp}&\equiv \frac{1}{2}\cbet{2}, & \cbetm[0^-]{\perp\perp}&\equiv0, \\
    \cbetm[1^+]{\perp\perp}&\equiv\frac{1}{6}(\cbet{1}+2\cbet{3}), & \cbetm[1^-]{\perp\perp}&\equiv\frac{1}{6}(2\cbet{1}+\cbet{2}), \\
    \cbetm[2^+]{\perp\perp}&\equiv\frac{1}{2}\cbet{1}, & \cbetm[2^-]{\perp\perp}&\equiv0,
  \end{aligned}
  \label{courel}
\end{equation}
where we find ${\cbetm[E]{\perp\parallel}\equiv\cbetm[E]{\parallel\perp}}$.
We can thus summarise some important relations for nonvanishing transfer couplings as
\begin{equation}
  \frac{\calpm[A]{\parallel\parallel}}{\calpm[A]{\perp\perp}}\equiv
  \frac{\calpm[A]{\perp\parallel}}{\calpm[A]{\parallel\perp}}\equiv
  \frac{\alpm[A]{\parallel\parallel}}{\alpm[A]{\perp\perp}}\equiv
  \frac{\alpm[A]{\perp\parallel}}{\alpm[A]{\parallel\perp}}=2, \quad
  \frac{\cbetm[E]{\perp\parallel}}{\cbetm[E]{\parallel\perp}}\equiv
  \frac{\betm[E]{\perp\parallel}}{\betm[E]{\parallel\perp}}=1,
  \label{doubleplusgood}
\end{equation}
with two sets of exceptions in the rotational sector
\begin{equation}
  \begin{gathered}
  \frac{\calpm[1^+]{\parallel\parallel}}{\calpm[1^+]{\perp\perp}}
  \equiv
  \frac{\calpm[1^+]{\perp\parallel}}{\calpm[1^+]{\parallel\perp}}
  \equiv
  \frac{\alpm[1^+]{\parallel\parallel}}{\alpm[1^+]{\perp\perp}}
  \equiv
  \frac{\alpm[1^+]{\perp\parallel}}{\alpm[1^+]{\parallel\perp}}=-2,\\
  \frac{\calpm[0^+]{\parallel\parallel}}{\calpm[0^+]{\perp\perp}}
  \equiv
  \frac{\calpm[0^+]{\parallel\perp}}{\calpm[0^+]{\perp\parallel}}
  \equiv
  \frac{\alpm[0^+]{\parallel\parallel}}{\alpm[0^+]{\perp\perp}}
  \equiv
  \frac{\alpm[0^+]{\parallel\perp}}{\alpm[0^+]{\perp\parallel}}=2.
  \label{plusgood}
  \end{gathered}
\end{equation}
The resulting effect of the multipliers in the Lagrangian picture~\cref{sysrie,sysriex} translates to
\begin{subequations}
  \begin{align}
    \calp{1}\neq 0&\Rightarrow \cR[\ovl{ij}]{A2p}+\ncR[\ovl{ij}]{A2p}
    \approx \cR[\ovl{ijk}]{A2m}-\ncR[\ovl{ijk}]{A2m}\approx 0,\label{ghoqor1}\\
    \calp{2}\neq 0&\Rightarrow \cR{A0m}+\ncR{A0m}\approx \cR[\ovl{ij}]{A1p}-\ncR[\ovl{ij}]{A1p}
    \nonumber\\
    &\ \ \ 
    \approx \cR[\ovl{ijk}]{A2m}+\ncR[\ovl{ijk}]{A2m}\approx 0,\label{ghoqor2}\\
    \calp{3}\neq 0&\Rightarrow \cR{A0m}-\ncR{A0m}\approx 0,\label{ghoqor3}\\
    \calp{4}\neq 0&\Rightarrow \cR{A0p}-2\ncR{A0p}\approx \cR[\ovl{i}]{A1m}+\ncR[\ovl{i}]{A1m}
    \nonumber\\
    &\ \ \ 
    \approx\cR[\ovl{ij}]{A2p}-\ncR[\ovl{ij}]{A2p} \approx 0,\label{ghoqor4}\\
    \calp{5}\neq 0&\Rightarrow \cR[\ovl{ij}]{A1p}+\ncR[\ovl{ij}]{A1p}\approx \cR[\ovl{i}]{A1m}-\ncR[\ovl{i}]{A1m}\approx 0,\label{ghoqor5}\\
    \calp{6}\neq 0&\Rightarrow \cR{A0p}+2\ncR{A0p}\approx 0\label{ghoqor6} \\
    \cbet{1}\neq 0&\Rightarrow \cT[\ovl{ij}]{B1p}-\ncT[\ovl{ij}]{B1p}\approx \cT[\ovl{i}]{B1m}-2\ncT[\ovl{i}]{B1m}
    \nonumber\\
    &\ \ \ 
   \approx \cT[\ovl{ijk}]{A2m}\approx \ncT[\ovl{ij}]{B2p}\approx 0,\label{ghoqor7}\\
    \cbet{2}\neq 0&\Rightarrow \cT[\ovl{i}]{B1m}+\ncT[\ovl{i}]{B1m}\approx \ncT{B0p}\approx 0,\label{ghoqor8}\\
    \cbet{3}\neq 0&\Rightarrow \cT[\ovl{ij}]{B1p}+2\ncT[\ovl{ij}]{B1p}\approx \cT{A0m}\approx 0.\label{ghoqor9}
  \end{align}
\end{subequations}

\section{Linearisation of sure primary first-class constraints}\label{linearisation_doubts}

In this appendix we will consider the safety of including the linearised sSFCs in the final d.o.f count. We recall that the Poincar\'e gauge symmetry implies the existence of 10 sSFCs, labelled $\tensor{\mathcal{  H}}{_\perp}$, $\tensor{\mathcal{  H}}{_\alpha}$, $\tensor{\mathcal{  H}}{_{\ovl{ij}}}$ and $\tensor{\mathcal{  H}}{_{\perp\ovl{i}}}$. However we frequently found in~\cite{chapter4} that some of these quantitites were missing when linearised on the PiC shell. An sSFC may clearly vanish if it is an arbitrary linear combination of iPFCs (PiCs which are FC in the final analysis), consistent with its FC property; how then to interpret an sSFC which happens to be an arbitrary linear combination of iSSCs?

This problem is resolved when we see that~\crefrange{total_hamiltonian}{total_hamiltonian_fin} are \emph{incomplete} formulae for the sSFCs when iPSCs (PiCs which are SC in the final analysis) are present in the theory. Let the super-Hamiltonian be, to lowest perturbative order, a linear combination of the only two iPSCs which appear in a given theory
\begin{equation}
  \tensor{\mathcal{  H}}{^\flat_\perp}\equiv
  \tensor{c}{_{A}^{\acu{u}}}\tensor[^A]{\varphi}{^\flat_{\acu{u}}}
  +\tensor{c}{_{E}^{\acu{u}}}\tensor[^E]{\varphi}{^\flat_{\acu{u}}}
  \approx 0,
  \label{spindof}
\end{equation}
and note that the only nonvanishing commutator between PiCs $\big\{\tensor[^A]{\varphi}{^\flat_{\acu{u}}},\tensor[^E]{\varphi}{^\flat_{\acu{u}}}\big\}$ will be of order unity.
The total Hamiltonian will take the form
\begin{equation}
  \begin{aligned}
  \tensor{\mathcal{  H}}{_{\text{T}}}&\equiv N\tensor{\mathcal{  H}}{^\flat_\perp}+
  \tensor[^A]{{u}}{^\flat^{\acu{u}}}\tensor[^A]{\varphi}{^\flat_{\acu{u}}}
  +\tensor[^E]{{u}}{^\flat^{\acu{u}}}\tensor[^E]{\varphi}{^\flat_{\acu{u}}}+\cdots\\
  &\equiv
  N\tensor{\ovl{\mathcal{  H}}}{^\flat_\perp}+\cdots,
  \label{modham}
\end{aligned}
\end{equation}
where the elipses in~\eqref{modham} include the remaining sSFCs, iPFCs and surface terms, and all higher-order terms. The \emph{modified} super-Hamiltonian is formed by solving for the PiC multipliers, and we have
\begin{equation}
  \begin{aligned}
  \tensor{\ovl{\mathcal{  H}}}{^\flat_\perp}&\equiv
  \tensor{\mathcal{  H}}{^\flat_\perp}
-\tensor{\smash{\Big(\Big\{ \tensor[^E]{\smash{\varphi}}{^\flat},\tensor[^A]{\smash{\varphi}}{^\flat} \Big\}^{-1}\Big)}}{^{\acu{v}}_{\acu{u}}}
    \\
  &
\times
\Big\{ \tensor[^E]{\smash{\varphi}}{^\flat^{\acu{u}}},\tensor{\smash{\mathcal{  H}}}{^\flat_\perp} \Big\}
  \tensor[^A]{\smash{\varphi}}{^\flat_{\acu{v}}}
  +(A\leftrightarrow E)
  \approx 0.
  \label{modcer}
  \end{aligned}
\end{equation}
The quantity defined in~\eqref{modcer} is the linearisation of the \emph{complete} sure secondary, and is FC by construction. Moreover, we can see by substituting from~\eqref{spindof} that even this complete quantity will vanish, with or without reference to the PiC shell. The argument can be generalised to arbitrarily many iPSCs, and to the remaining sSFCs.

\section{Traditional, linearised simple spin $1^+$ case}\label{simpleA1p}

It is useful to analyse the Hamiltonian structure of the $1^+$ case without multipliers, as it was originally considered in~\cite{2002IJMPD..11..747Y}. Our findings will also be corroborated by the \emph{HiGGS} computer algebra software in~\cite{crunchy}. We note that the defining conditions in~\eqref{simspione} are consistent with curvature-free constraints: the PiCs depend only on the momenta. The nonlinear commutators of this theory are
\begin{subequations}
  \begin{align}
    \Big\{\pic{B0p},\pic{A0p}\Big\}&\approx -\frac{6\alp{0}}{J}\planck^2\delta^3,\label{B0pA0p}\\
    \Big\{\pic[\ovl{i}]{B1m},\pic[\ovl{l}]{B1m}\Big\}&\approx \frac{2}{J^2}\PiP[\ovl{il}]{B1p}\delta^3,\label{B1mB1m}\\
    \Big\{\pic[\ovl{i}]{B1m},\pic{A0p}\Big\}&\approx -\frac{1}{J^2}\PiP[\ovl{i}]{A1m}\delta^3,\label{B1mA0p}\\
    \Big\{\pic[\ovl{i}]{B1m},\pic{A0m}\Big\}&\approx -\frac{2}{J^2}\etau{\ovl{lp}}\etau{\ovl{mq}}\epsd{\ovl{ilm}}\PiP[\ovl{pq}]{A1p}\delta^3,\label{B1mA0m}\\
    \Big\{\pic[\ovl{i}]{B1m},\pic[\ovl{lm}]{A2p}\Big\}&\approx \frac{1}{2J^2}\etad{\ovl{i}\langle\ovl{l}}\PiP[\ovl{m}\rangle]{A1m}\delta^3,\label{B1mA2p}\\
    \Big\{\pic[\ovl{i}]{B1m},\pic[\ovl{lmn}]{A2m}\Big\}&\approx \frac{1}{2J^2}\Big(
	\etad{\ovl{in}}\PiP[\ovl{lm}]{A1p}
	+\etad{\ovl{i}{[}\ovl{m}}\PiP[\ovl{n}{]}\ovl{l}]{A1p}
	\nonumber\\
	&
	\quad
	\quad
	\quad
	\quad
	+\frac{3}{2}\etad{\ovl{n}{[}\ovl{m}}\PiP[\ovl{l}{]}\ovl{i}]{A1p}
      \Big)\delta^3,\label{B1mA2m}\\
      \Big\{\pic[\ovl{ij}]{B2p},\pic[\ovl{lm}]{B2p}\Big\}&\approx -\frac{2}{J^2}\etad{(\ovl{i}|(\ovl{l}}\PiP[\ovl{m})|\ovl{j})]{B1p}\delta^3,\label{B2pB2p}\\
  \left\{
\pic[\ovl{ij}]{B2p}
,
\pic[\ovl{lm}]{A2p}
  \right\}&\approx
  \Big(
  \frac{\alp{0}\planck^2}{J}\etad{\ovl{i}(\ovl{l}}\etad{\ovl{m})\ovl{j}}
  \nonumber\\
  &\quad
  \quad
  \quad
  \quad
  \quad
  +\frac{1}{J^2}\etad{\langle\ovl{i}|\langle\ovl{l}}\PiP[\ovl{m}\rangle |\ovl{j}\rangle]{A1p}
\Big)
\delta^3,
\label{B2pA2p}
\\
      \Big\{\pic[\ovl{ij}]{B2p},\pic[\ovl{lmn}]{A2m}\Big\}&\approx -\frac{1}{J^2}\Proj[_{\ovl{lmn}}^{\ovl{pqr}}]{A2m}\etad{\ovl{r}\langle\ovl{i}}\etad{\ovl{j}\rangle\ovl{p}}\PiP[\ovl{q}]{A1m}\delta^3.
      \label{B2pA2m}
  \end{align}
\end{subequations}
On the PiC shell, the linearised sSFCs of the minimal theory are
  \begin{align}
    \haml{mom0p}&\approx
    \frac{1}{2}\alp{0}\planck^2J\cRl{A0p}
    \approx 0,
    \\
    \haml[\ovl{i}]{mom1m}&\approx
    -\etaul{\ovl{jk}}\covderl{\ovl{j}}\PiPl[\ovl{ik}]{B1p}-\alp{0}\planck^2J\cRl[\ovl{i}]{A1m}
      \approx 0,\\
    \haml[\ovl{ij}]{rot1p}&\approx
    \covderl{{[}\ovl{i}}\PiPl[\ovl{j}{]}]{A1m}+2\PiPl[\ovl{ij}]{B1p}-\alp{0}\planck^2J\cTl[\ovl{ij}]{B1p}
    \approx 0,
    \\
    \haml[\ovl{i}]{rot1m}&\approx
      \etaul{\ovl{jk}}\covderl{\ovl{j}}\PiPl[\ovl{ik}]{A1p}
      -\alp{0}\planck^2\Jl\cTl[\ovl{i}]{B1m}
      \approx 0.
      \label{cerled}
  \end{align}
The Einstein--Hilbert term contributes independent parts of the Riemann--Cartan and torsion tensors to each irrep equation, thus subtracting $2\times 10$ canonical d.o.f even at the linear level.

Some of the commutators in~\crefrange{B0pA0p}{B2pA2m} also survive at the linear level, so that we do not have to worry about the consistency conditions of $\picl{B0p}$, $\picl{B2p}$, $\picl{A0p}$ and $\picl{A2p}$. The consistencies of $\picl{B1m}$ and $\picl{A0m}$ suggest the following secondaries on the combined shell of PiCs and sSFCs
\begin{subequations}
  \begin{align}
    \sicl[\ovl{i}]{B1m}&\approx
    -\frac{2}{\Jl}\etaul{jk}\covderl{\ovl{j}}\PiPl[\ovl{ik}]{B1p}-\frac{\alp{0}\planck^2}{5\alp{5}\Jl}\PiPl[\ovl{i}]{A1m}+\alp{0}\planck^2\cRl[\ovl{i}]{A1m}
    \nonumber\\
    &\approx 0,
    \\
    \Big\{&\sicl[\ovl{i}]{B1m},\picl[\ovl{l}]{B1m}\Big\}\approx
    -\frac{\alp{0}^2\planck^2}{2\alp{5}\Jl}\etadl{\ovl{il}}\delta^3,
    \\
    \sicl{A0m}&\approx
    -\frac{2}{\Jl}\epsul{\ovl{ijk}}\covderl{\ovl{i}}\PiPl[\ovl{jk}]{A1p}-(\alp{0}-8\bet{3})\planck^2\cTl{A0m}
    \nonumber\\
    &
    \approx 0,
    \\
    \Big\{&\sicl{A0m},\picl{A0m}\Big\}\approx
    -\frac{24\planck^2(\alp{0}-8\bet{3})}{\Jl}\delta^3,
  \end{align}
\end{subequations}
where these SiCs are SC at $\mathcal{  O}(1)$, since they fail to commute with the PiCs which invoke them. This is to be expected from the theory of conjugate pairs~\cite{1983PhRvD..28.2455B}. We will not obtain the secondary $\sicl{A2m}$ deriving from $\picl{A2m}$, since the square of the tensor part projection entails calculations which are difficult on paper, however we note that these quantities should also form a conjugate SC pair.

In the final counting therefore, all the PiCs and SiCs are SC, and the linearised theory propagates a total of \emph{five} d.o.f
\begin{equation}
\begin{aligned}
  5&=\smash{\frac{1}{2}}\big(80-2\times 10[\text{sPFC}]-2\times 10[\text{sSFC}]
    \\
    &\ \ \ \ \ \ \ 
    -(1+3+5+1+1+5+5)[\text{iPSC}]\\
    &\ \ \ \ \ \ \ 
    -(3+1+5)[\text{iSSC}]\big).
    \label{<+label+>}
\end{aligned}
\end{equation}
These d.o.f are interpreted as the massless graviton and a massive vector mode, so that the findings of~\cite{2002IJMPD..11..747Y} are confirmed. As discussed in~\cref{lorspe} and shown explicitly in~\cite{2002IJMPD..11..747Y}, consideration of the nonlinear commutators brings this count to eight d.o.f.
\section{The surficial commutator}\label{surfic}
An ostensibly limiting factor in previous Hamiltonian analyses of the PGT~\cite{1983PhRvD..28.2455B,2002IJMPD..11..747Y,1999IJMPD...8..459Y} is the dependence of various commutators on the \emph{spatial gradient} of the equal-time Dirac function. The coefficients of such gradients are generally gauge-dependent, while standard texts~\cite{Henneaux:1992ig,blagojevic2002gravitation} do not (to our knowledge) provide a prescription for their covariant interpretation (see, however, excellent discussions of special cases in electrodynamics~\cite{blagrec1} and noncritical string theory~\cite{blagrec2}).
In this appendix we provide the covariant extension to the `Poisson bracket formula', which eliminates these gradients through the use of surface terms. The resulting expression is more costly to evaluate than the original by a factor of only several, allowing us to proceed farther into the theory.

Our starting point is the realisation that the Poisson bracket is ultimately motivated by the time derivative operator. We consider the time derivative of the covariant quantity $\fA$, which is assumed to depend canonically on a collection of (matter or gravitational) fields $\{\fphi\}$ and their conjugate momenta $\{\fpi\}$, along with their \emph{first} covariant derivatives $\{\coder{\mu}\fphi\}$ and $\{\coder{\mu}\fpi\}$. The total Hamiltonian is assumed to contain a term bilinear in two further covariant quantities ${\mathcal{  H}_{\text{T}}\supset \fB\fC}$, and so a simple algebra reveals that the velocity $\tensor{\dot{\mathcal{  A}}}{_{\acu{u}}}$ contains terms of the form
\begin{widetext}
  \begin{align}
    \tensor{\dot{\mathcal{  A}}}{_{\acu{u}}}(x_1)&\supset
    \int\mathrm{d}^3x_2\big\{\fA(x_1) ,\fB(x_2) \big\}\fC(x_2)\equiv
    \Bigg[
    \left( 
      \covard{\fA}{\fphi}\cdot\covard{\fB}{\fpi}
    -\covard{\fA}{\fpi}\cdot\covard{\fB}{\fphi} 
     \right)\fC
     \nonumber\\
    +&\coder{\alpha}\bigg[
    \left(
      \pard{\fA}{\coder{\alpha}\fphi}\cdot\covard{\fB}{\fpi}
      -\pard{\fA}{\coder{\alpha}\fpi}\cdot\covard{\fB}{\fphi}
    \right) \fC 
    \bigg]
    +\left(
      \pard{\fB}{\coder{\alpha}\fphi}\cdot\covard{\fA}{\fpi}
      -\pard{\fB}{\coder{\alpha}\fpi}\cdot\covard{\fA}{\fphi}
    \right) \coder{\alpha}\fC
    \nonumber\\
    -&\coder{\alpha}\bigg[
    \bigg(
      \pard{\fA}{\coder{\alpha}\fphi}\cdot\pard{\fB}{\coder{\beta}\fpi}
      -\pard{\fA}{\coder{\alpha}\fpi}\cdot\pard{\fB}{\coder{\beta}\fphi}
    \bigg) \coder{\beta}\fC 
  \bigg]
\Bigg]\Bigg\rvert_{x_1},
  \label{simexp}
  \end{align}
\end{widetext}
where the dot product sums over field species and we construct a derivative which naturally extends the variational derivative on a scalar Lagrangian to tensors of arbitrary rank
\begin{equation}
  \covard{\fA}{\fphi}\equiv \copard{\fA}{\fphi}-\coder{\alpha}\left( \pard{\fA}{\coder{\alpha}\fphi} \right).
  \label{bar_derivative}
\end{equation}

In~\eqref{bar_derivative} the notation ${\bar{\partial}/\bar{\partial}\fphi}$ indicates that $\coder{\alpha}\fphi$ is held constant when evaluating the partial derivative~\cite{blagojevic2002gravitation}. 
It is only expressions such as~\eqref{simexp} which must be covariant, and the operations ${\bar{\partial}/\bar{\partial}\fphi}$, ${\partial/\partial\coder{\alpha}\fphi}$, ${\bar{\delta}/\bar{\delta}\fphi}$ and their momentum counterparts all support that property. Therefore, we find it most natural to express the Poisson bracket as the \emph{kernel} which reproduces~\eqref{simexp}. In general, this kernel takes the form of the second-order covariant differential operator 
\begin{widetext}
  \begin{align}
	  \big\{\fA(x_1),\fB(x_2) \big\}&\equiv
    \bigg[
      \copard{\fA}{\fphi}\cdot\covard{\fB}{\fpi}
      -\copard{\fA}{\fpi}\cdot\covard{\fB}{\fphi}
      +\pard{\fA}{\coder{\alpha}\fphi}\cdot\coder{\alpha}\bigg(\covard{\fB}{\fpi}\bigg)
      -\pard{\fA}{\coder{\alpha}\fpi}\cdot\coder{\alpha}\bigg(\covard{\fB}{\fphi}\bigg)
    \bigg]\delta^3
    \nonumber\\
    &
    \ \ +\bigg[
      \pard{\fA}{\coder{\alpha}\fphi}\cdot\covard{\fB}{\fpi}
      -\pard{\fA}{\coder{\alpha}\fpi}\cdot\covard{\fB}{\fphi}
      +\copard{\fA}{\fpi}\cdot\pard{\fB}{\coder{\alpha}\fphi}
      -\copard{\fA}{\fphi}\cdot\pard{\fB}{\coder{\alpha}\fpi}
      \nonumber\\
      &
      \hphantom{\ \ \ \ \ \ \ \ \ \ \ \ \  }
      +\pard{\fA}{\coder{\beta}\fpi}\cdot\coder{\beta}\bigg(\pard{\fB}{\coder{\alpha}\fphi}\bigg)
      -\pard{\fA}{\coder{\beta}\fphi}\cdot\coder{\beta}\bigg(\pard{\fB}{\coder{\alpha}\fpi}\bigg)
    \bigg]\delta^3\coder{\alpha}
    \nonumber\\
    &
    \ \ +
    \bigg(
      \pard{\fA}{\coder{\alpha}\fpi}\cdot\pard{\fB}{\coder{\beta}\fphi}
      -\pard{\fA}{\coder{\alpha}\fphi}\cdot\pard{\fB}{\coder{\beta}\fpi}
    \bigg) \delta^3\coder{\alpha}\coder{\beta}.
  \label{defpoi}
  \end{align}
\end{widetext}
This concludes our discussion of the surficial commutator for the first-order Euler--Lagrange formalism.
\end{document}

%% file: macros.tex
\newrobustcmd{\Mgra}[1]{%
  {\tensor{M}{_{\text{#1}}}}%
}
\newrobustcmd{\Mpro}[1]{%
  {\tensor{\mathproper{M}}{_{\text{#1}}}}%
}
\newrobustcmd{\Malt}[1]{%
  {\tensor{\mathscr{M}}{_{\text{#1}}}}%
}
\newrobustcmd{\Mkom}[1]{%
  {\tensor{\mathfrak{M}}{_{\text{#1}}}}%
}

\newrobustcmd{\Mtotal}{%
  {\tensor{M}{_{\text{T}}}}%
}

\newrobustcmd{\Qtotal}{%
  {\tensor{Q}{_{\text{T}}}}%
}

\newrobustcmd{\Qtotalcal}{%
  {\tensor{\mathcal{  Q}}{_{\text{T}}}}%
}

\newrobustcmd{\action}[1]{%
  {\tensor{S}{_{\text{#1}}}}%
}

\newrobustcmd{\lagrangian}[1]{%
  {\tensor{L}{_{\text{#1}}}}%
}

\newrobustcmd{\lagrangianprop}[1]{%
  {\tensor{\mathproper{L}}{_{\text{#1}}}}%
}

\newrobustcmd{\epl}{%
  {\tensor{\mathsf{e}}{_+}}%
}

\newrobustcmd{\epe}{%
  {\tensor{\mathsf{e}}{_\perp}}%
}

\newrobustcmd{\qz}{%
  {\text{\color{orange}\cmark}}%
}
\newrobustcmd{\jz}{%
  {\text{\color{red}\xmark}}%
}

\newrobustcmd{\projmatrix}[2][placeholder]{%
  {\tensor*{M}{_{#1}^{#2}}}
}
\newrobustcmd{\projorthhum}[2][placeholder]{%
  {\tensor[^#2]{\smash{\check{\mathcal{  P}}}}{#1}}
}
\newrobustcmd{\projorthhumu}[2][placeholder]{%
  {\tensor[^#2]{\smash{\check{\mathcal{  P}}}}{#1}}
}
\newrobustcmd{\projorth}[2][placeholder]{%
  {\tensor[^#2]{\smash{\hat{\mathcal{  P}}}}{#1}}
}
\newrobustcmd{\projlore}[2][placeholder]{%
  {\tensor[^#2]{\hat{\mathcal{  P}}}{#1}}
}

\newrobustcmd{\gensec}[3][placeholder]{%
  {\tensor*[^#1]{\chi}{^{#2}_{\acu{#3}}}}
}

\newrobustcmd{\glfourr}{%
  {\mathrm{GL}(4,\mathbb{R})}%
}

\newrobustcmd{\sltwoc}{%
  {\mathrm{SL}(2,\mathbb{C})}%
}

\newrobustcmd{\poincare}{%
  {\mathbb{R}^{1,3}\rtimes\mathrm{SO}^+(1,3)}%
}

\newrobustcmd{\poincaref}{%
  {\mathrm{P}(1,3)}%
}

\newrobustcmd{\weyl}{%
  {\mathrm{W}(1,3)}%
}

\newrobustcmd{\conformal}{%
  {\mathrm{C}(1,3)}%
}

\newrobustcmd{\diffeomorphism}{%
  {\mathbb{R}^{1,3}}%
}

\newrobustcmd{\soonethree}{%
  {\mathrm{SO}^+(1,3)}%
}

\newrobustcmd{\othree}{%
  {\mathrm{SO}(3)}%
}

\newrobustcmd{\sothree}{%
  {\mathrm{SO}(3)}%
}

\newrobustcmd{\sotwo}{%
  {\mathrm{SO}(2)}%
}

\newrobustcmd{\suthreec}{%
  {\mathrm{SU}(3)_{\text{c}}}%
}

\newrobustcmd{\sutwol}{%
  {\mathrm{SU}(2)_{\text{L}}}%
}

\newrobustcmd{\uoney}{%
  {\mathrm{U}(1)_{\text{Y}}}%
}

\newrobustcmd{\uone}{%
  {\mathrm{U}(1)}%
}

\newrobustcmd{\uoneem}{%
  {\mathrm{U}(1)_{\text{em}}}%
}

\newrobustcmd{\sutwo}{%
  {\mathrm{SU}(2)}%
}

\newrobustcmd{\eplus}{%
  {\tensor{\mathsf{e}}{_{+}}}%
}

\newrobustcmd{\esf}[1]{%
  {\tensor{\mathsf{e}}{_{#1}}}
}%
\newrobustcmd{\esfu}[1]{%
  {\tensor{\mathsf{e}}{^{#1}}}
}%
\newrobustcmd{\gam}[1]{%
  {\tensor{\gamma}{_{#1}}}
}%
\newrobustcmd{\gamu}[1]{%
  {\tensor{\gamma}{^{#1}}}
}%

\newrobustcmd{\planck}{%
  {m_{\text{p}}}%
}

\newrobustcmd{\caligR}{%
  {\mathcal{R}}%
}
\newrobustcmd{\caligT}{%
  {\mathcal{T}}%
}

\newrobustcmd{\pgt}{%
  PGT\textsuperscript{q,+}\ %
}

\newrobustcmd{\unl}[1]{%
  {\mathfrak{#1}}%
}
\newrobustcmd{\ovl}[1]{%
\overline{#1}%
}
\newrobustcmd{\acu}[1]{%
\acute{#1}%
}

\newrobustcmd{\indiq}[2][placeholder]{%
\IfEqCase{#1}{%
  {placeholder}{%
    \IfEqCase{#2}{%
      {1}{\ovl{k}}%
      {2}{\ovl{kl}}%
      {3}{\ovl{klm}}%
    }%
  }%
}[#1]%
}%

\newrobustcmd{\indaq}[2][placeholder]{%
\IfEqCase{#1}{%
  {placeholder}{%
    \IfEqCase{#2}{%
      {1}{\overline{k}}%
      {2}{\overline{kl}}%
      {3}{\overline{klm}}%
    }%
  }%
}[#1]%
}%

\newrobustcmd{\indeq}[2][placeholder]{%
\IfEqCase{#1}{%
  {placeholder}{%
    \IfEqCase{#2}{%
      {1}{k}%
      {2}{kl}%
      {3}{klm}%
    }%
  }%
}[#1]%
}%

\newrobustcmd{\indoq}[2][placeholder]{%
\IfEqCase{#1}{%
  {placeholder}{%
    \IfEqCase{#2}{%
      {1}{\alpha}%
      {2}{\alpha\beta}%
      {3}{\alpha\beta\gamma}%
    }%
  }%
}[#1]%
}%

\newrobustcmd{\fcphi}[1]{%
  \tensor[^{\text{FC}}]{\phi}{_{#1}}%
}

\newrobustcmd{\scphi}[1]{%
  \tensor[^{\text{SC}}]{\phi}{_{#1}}%
}

\newrobustcmd{\fcmul}[1]{%
  \tensor[^{\text{FC}}]{\upsilon}{_{#1}}%
}

\newrobustcmd{\arb}{%
  {\tensor{f}{_{\text{lin}}}}%
}

\newrobustcmd{\scmul}[1]{%
  \tensor[^{\text{SC}}]{\upsilon}{_{#1}}%
}

\newrobustcmd{\foli}[1]{%
\tensor{n}{_{#1}}%
}

\newrobustcmd{\foliu}[1]{%
\tensor{n}{^{#1}}%
}

\newrobustcmd{\covderl}[1]{%
\tensor{\mathcal{D}}{^{\flat}_{\indiq[#1]{1}}}%
}

\newrobustcmd{\covder}[1]{%
\tensor{\mathcal{D}}{_{\indiq[#1]{1}}}%
}
\newrobustcmd{\coder}[1]{%
\tensor{D}{_{\indiq[#1]{1}}}%
}

\newrobustcmd{\deltal}[2]{%
  \tensor*{\delta}{_{\phantom{\flat}}^{\flat}_{#1}^{#2}}%
}

\newrobustcmd{\deltaud}[2]{%
  \tensor*{\delta}{^{#1}_{#2}}%
}
\newrobustcmd{\etau}[1]{%
\tensor{\eta}{^{\indiq[#1]{2}}}%
}
\newrobustcmd{\etaul}[1]{%
\tensor{\eta}{^{\flat}^{\indiq[#1]{2}}}%
}
\newrobustcmd{\etad}[1]{%
\tensor{\eta}{_{\indiq[#1]{2}}}%
}
\newrobustcmd{\etadl}[1]{%
\tensor{\eta}{^{\flat}_{\indiq[#1]{2}}}%
}

\newrobustcmd{\epsul}[1]{%
\tensor{\epsilon}{^{\flat}^{\indiq[#1]{3}}^{\perp}}
}
\newrobustcmd{\epsdl}[1]{%
\tensor{\epsilon}{^{\flat}_{\indiq[#1]{3}}_{\perp}}
}
\newrobustcmd{\epsd}[1]{%
\tensor{\epsilon}{_{\indiq[#1]{3}}_{\perp}}
}
\newrobustcmd{\epsu}[1]{%
\tensor{\epsilon}{^{\indiq[#1]{3}}^{\perp}}
}

\newrobustcmd{\hfl}[2]{%
  \tensor{h}{^{\flat}_{#1}^{#2}}
}

\newrobustcmd{\bet}[1]{%
  \tensor{\hat{\beta}}{_{#1}}
}
\newrobustcmd{\alp}[1]{%
  \tensor{\hat{\alpha}}{_{#1}}
}
\newrobustcmd{\cgalp}{\tensor{\alpha}{_{\text{CG}}}}

\newrobustcmd{\cbet}[1]{%
  \tensor{\bar{\beta}}{_{#1}}
}
\newrobustcmd{\calp}[1]{%
  \tensor{\bar{\alpha}}{_{#1}}
}

\newrobustcmd{\alpg}[1]{%
  \tensor{\check{\alpha}}{_{#1}}
}
\newrobustcmd{\betg}[1]{%
  \tensor{\check{\beta}}{_{#1}}
}
\newrobustcmd{\calpg}[1]{%
  \tensor{\acu{\alpha}}{_{#1}}
}
\newrobustcmd{\cbetg}[1]{%
  \tensor{\acu{\beta}}{_{#1}}
}

\newrobustcmd{\hub}{%
  {\underline{\mathsf{h}}}
}
\newrobustcmd{\hubm}{%
  {\underline{\mathsf{h}}^{-1}}
}
\newrobustcmd{\hob}{%
  {\bar{\mathsf{h}}}
}
\newrobustcmd{\hobm}{%
  {\bar{\mathsf{h}}^{-1}}
}
\newrobustcmd{\hdet}{%
  {\det \mathsf{h}}
}
\newrobustcmd{\hmdet}{%
  {\det \mathsf{h}^{-1}}
}
\newrobustcmd{\Rsf}{%
  {\mathsf{R}}
}

\newrobustcmd{\alpm}[2][placeholder]{%
  {\tensor*{\hat{\alpha}}{_{#1}^{#2}}}
}
\newrobustcmd{\calpm}[2][placeholder]{%
  \tensor*{\bar{\alpha}}{_{#1}^{#2}}
}

\newrobustcmd{\betm}[2][placeholder]{%
  {\tensor*{\hat{\beta}}{_{#1}^{#2}}}
}
\newrobustcmd{\cbetm}[2][placeholder]{%
  \tensor*{\bar{\beta}}{_{#1}^{#2}}
}

\newrobustcmd{\lamr}{%
  {\zeta_{\mathcal{  R}} }
}
\newrobustcmd{\barlamr}{%
  {\bar{\zeta}_{\mathcal{  R}} }
}
\newrobustcmd{\lamt}{%
  {\zeta_{\mathcal{  T}} }
}
\newrobustcmd{\barlamt}{%
  {\bar{\zeta}_{\mathcal{  T}} }
}

\newrobustcmd{\atmp}[1]{%
  \tensor{\hat{a}}{_{#1}}
}
\newrobustcmd{\btmp}[1]{%
  \tensor{b}{_{#1}}
}

\newrobustcmd{\ctmp}[2][placeholder]{%
  {\tensor*{c}{_{#1}^{#2}}}
}
\newrobustcmd{\dtmp}[2][placeholder]{%
  {\tensor*{d}{_{#1}^{#2}}}
}

\newrobustcmd{\etmp}[1]{%
  \tensor{e}{_{#1}}
}

\newrobustcmd{\batmp}[1]{%
  \tensor{\ovl{a}}{_{#1}}
}
\newrobustcmd{\bbtmp}[1]{%
  \tensor{\ovl{b}}{_{#1}}
}
\newrobustcmd{\bctmp}[1]{%
  \tensor{\ovl{c}}{_{#1}}
}
\newrobustcmd{\bdtmp}[1]{%
  \tensor{\ovl{d}}{_{#1}}
}
\newrobustcmd{\betmp}[1]{%
  \tensor{\ovl{e}}{_{#1}}
}

\newrobustcmd{\ptl}[1]{%
  \tensor{\partial}{#1}
}

\newrobustcmd{\etaf}[1]{%
  \tensor{\eta}{#1}
}

\newrobustcmd{\epsf}[1]{%
  \tensor{\epsilon}{#1}
}

\newrobustcmd{\RSO}[2][placeholder]{%
  {\tensor[^{#2}]{\mathcal{  R}}{#1}}
}
\newrobustcmd{\TSO}[2][placeholder]{%
  {\tensor[^{#2}]{\mathcal{  T}}{#1}}
}
\newrobustcmd{\FSO}[2][placeholder]{%
  {\tensor[^{#2}]{\mathcal{  F}}{#1}}
}
\newrobustcmd{\spinSO}[2][placeholder]{%
  {\tensor[^{#2}]{\sigma}{#1}}
}
\newrobustcmd{\RLambdaSO}[2][placeholder]{%
  {\tensor[^{#2}]{\zeta}{#1}}
}
\newrobustcmd{\TLambdaSO}[2][placeholder]{%
  {\tensor[^{#2}]{\zeta}{#1}}
}
\newrobustcmd{\KSO}[2][placeholder]{%
  {\tensor[^{#2}]{\mathcal{  K}}{#1}}
}

\newrobustcmd{\bper}[2][placeholder]{%
\IfEqCase{#2}{%
  {s}{\tensor{\mathfrak{s}}{#1}}%
  {a}{\tensor{\mathfrak{a}}{#1}}%
  {sbar}{\tensor{\bar{\mathfrak{s}}}{#1}}%
}[\packageError{cosmicclass}{Unidentified Critical Case: #1}{}]%
}

\newrobustcmd{\Jl}{%
  {J^{\flat}}%
}%

\newrobustcmd{\Nl}{%
  {N^{\flat}}%
}%

\newrobustcmd{\haml}[2][placeholder]{%
\IfEqCase{#2}{%
{mom0p}{\tensor{\mathcal{H}}{^{\flat}_{\perp}}}%
{mom1m}{\tensor{\mathcal{H}}{^{\flat}_{\indoq[#1]{1}}}}%
{rot1p}{\tensor{\mathcal{H}}{^{\flat}_{\indaq[#1]{2}}}}%
{rot1m}{\tensor{\mathcal{H}}{^{\flat}_{\perp}_{\indaq[#1]{1}}}}%
}[\packageError{cosmicclass}{Unidentified Critical Case: #1}{}]%
}

\newrobustcmd{\arc}[2][placeholder]{%
\IfEqCase{#2}{%
{B1p}{\tensor{\vartheta}{_{\perp\indiq[#1]{2}}}}%
{B2m}{\tensor[^{\text{T}}]{\vartheta}{_{\indiq[#1]{3}}}}%
{A0m}{\tensor[^{\text{P}}]{\vartheta}{}}%
{A1p}{\tensor{\overset{\wedge}{\vartheta}}{_{\perp\indiq[#1]{2}}}}%
{A1m}{\tensor{\overset{\rightharpoonup}{\vartheta}}{_{\indiq[#1]{1}}}}%
{A2p}{\tensor{\overset{\sim}{\vartheta}}{_{\perp\indiq[#1]{2}}}}%
{A2m}{\tensor[^{\text{T}}]{\vartheta}{_{\perp\indiq[#1]{3}}}}%
}[\packageError{cosmicclass}{Unidentified Critical Case: #1}{}]%
}

\newrobustcmd{\pic}[2][placeholder]{%
\IfEqCase{#2}{%
{B0p}{\varphi}%
{B1p}{\tensor{\overset{\wedge}{\varphi}}{_{\indiq[#1]{2}}}}%
{B1m}{\tensor{\varphi}{_{\perp\indiq[#1]{1}}}}%
{B2p}{\tensor{\overset{\sim}{\varphi}}{_{\indiq[#1]{2}}}}%
{A0p}{\tensor{\varphi}{_\perp}}%
{A0m}{\tensor[^{\text{P}}]{\varphi}{}}%
{A1p}{\tensor{\overset{\wedge}{\varphi}}{_{\perp\indiq[#1]{2}}}}%
{A1m}{\tensor{\overset{\rightharpoonup}{\varphi}}{_{\indiq[#1]{1}}}}%
{A2p}{\tensor{\overset{\sim}{\varphi}}{_{\perp\indiq[#1]{2}}}}%
{A2m}{\tensor[^{\text{T}}]{\varphi}{_{\indiq[#1]{3}}}}%
}[\packageError{cosmicclass}{Unidentified Critical Case: #1}{}]%
}

\newrobustcmd{\picu}[2][placeholder]{%
\IfEqCase{#2}{%
{B0p}{\varphi}%
{B1p}{\tensor{\smash{\overset{\wedge}{\varphi}}}{^{\indiq[#1]{2}}}}%
{B1m}{\tensor{\varphi}{^{\perp\indiq[#1]{1}}}}%
{B2p}{\tensor{\smash{\overset{\sim}{\varphi}}}{^{\indiq[#1]{2}}}}%
{A0p}{\tensor{\varphi}{_\perp}}%
{A0m}{\tensor[^{\text{P}}]{\varphi}{}}%
{A1p}{\tensor{\smash{\overset{\wedge}{\varphi}}}{^{\perp\indiq[#1]{2}}}}%
{A1m}{\tensor{\smash{\overset{\rightharpoonup}{\varphi}}}{^{\indiq[#1]{1}}}}%
{A2p}{\tensor{\smash{\overset{\sim}{\varphi}}}{^{\perp\indiq[#1]{2}}}}%
{A2m}{\tensor[^{\text{T}}]{\varphi}{^{\indiq[#1]{3}}}}%
}[\packageError{cosmicclass}{Unidentified Critical Case: #1}{}]%
}

\newrobustcmd{\picl}[2][placeholder]{%
\IfEqCase{#2}{%
{B0p}{\tensor{\varphi}{^{\flat}}}%
{B1p}{\tensor{\smash{\overset{\wedge}{\varphi}}}{^{\flat}_{\indiq[#1]{2}}}}%
{B1m}{\tensor{\varphi}{^{\flat}_{\perp}_{\indiq[#1]{1}}}}%
{B2p}{\tensor{\smash{\overset{\sim}{\varphi}}}{^{\flat}_{\indiq[#1]{2}}}}%
{A0p}{\tensor{\varphi}{_\perp}^{\flat}}%
{A0m}{\tensor[^{\text{P}}]{\varphi}{^{\flat}}}%
{A1p}{\tensor{\smash{\overset{\wedge}{\varphi}}}{^{\flat}_{\perp\indiq[#1]{2}}}}%
{A1m}{\tensor{\smash{\overset{\rightharpoonup}{\varphi}}}{^{\flat}_{\indiq[#1]{1}}}}%
{A2p}{\tensor{\smash{\overset{\sim}{\varphi}}}{^{\flat}_{\perp\indiq[#1]{2}}}}%
{A2m}{\tensor[^{\text{T}}]{\varphi}{^{\flat}_{\indiq[#1]{3}}}}%
}[\packageError{cosmicclass}{Unidentified Critical Case: #1}{}]%
}

\newrobustcmd{\mull}[2][placeholder]{%
\IfEqCase{#2}{%
{B0p}{\tensor{u}{^{\flat}}}%
{B1p}{\tensor{\smash{\overset{\wedge}{u}}}{^{\flat}_{\indiq[#1]{2}}}}%
{B1m}{\tensor{u}{^{\flat}_{\perp}_{\indiq[#1]{1}}}}%
{B2p}{\tensor{\smash{\overset{\sim}{u}}}{^{\flat}_{\indiq[#1]{2}}}}%
{A0p}{\tensor{u}{_\perp}^{\flat}}%
{A0m}{\tensor[^{\text{P}}]{u}{^{\flat}}}%
{A1p}{\tensor{\smash{\overset{\wedge}{u}}}{^{\flat}_{\perp\indiq[#1]{2}}}}%
{A1m}{\tensor{\smash{\overset{\rightharpoonup}{u}}}{^{\flat}_{\indiq[#1]{1}}}}%
{A2p}{\tensor{\smash{\overset{\sim}{u}}}{^{\flat}_{\perp\indiq[#1]{2}}}}%
{A2m}{\tensor[^{\text{T}}]{u}{^{\flat}_{\indiq[#1]{3}}}}%
}[\packageError{cosmicclass}{Unidentified Critical Case: #1}{}]%
}

\newrobustcmd{\mul}[2][placeholder]{%
\IfEqCase{#2}{%
{B0p}{\tensor{u}{}}%
{B1p}{\tensor{\smash{\overset{\wedge}{u}}}{_{\indiq[#1]{2}}}}%
{B1m}{\tensor{u}{_{\perp}_{\indiq[#1]{1}}}}%
{B2p}{\tensor{\smash{\overset{\sim}{u}}}{_{\indiq[#1]{2}}}}%
{A0p}{\tensor{u}{_\perp}}%
{A0m}{\tensor[^{\text{P}}]{u}{}}%
{A1p}{\tensor{\smash{\overset{\wedge}{u}}}{_{\perp\indiq[#1]{2}}}}%
{A1m}{\tensor{\smash{\overset{\rightharpoonup}{u}}}{_{\indiq[#1]{1}}}}%
{A2p}{\tensor{\smash{\overset{\sim}{u}}}{_{\perp\indiq[#1]{2}}}}%
{A2m}{\tensor[^{\text{T}}]{u}{_{\indiq[#1]{3}}}}%
}[\packageError{cosmicclass}{Unidentified Critical Case: #1}{}]%
}

\newrobustcmd{\PiP}[2][placeholder]{%
\IfEqCase{#2}{%
{B0p}{\hat{\pi}}%
{B1p}{\tensor{\overset{\wedge}{\hat{\pi}}}{_{\indiq[#1]{2}}}}%
{B1m}{\tensor{\hat{\pi}}{_{\perp\indiq[#1]{1}}}}%
{B2p}{\tensor{\overset{\sim}{\hat{\pi}}}{_{\indiq[#1]{2}}}}%
{A0p}{\tensor{\hat{\pi}}{_\perp}}%
{A0m}{\tensor[^{\text{P}}]{\hat{\pi}}{}}%
{A1p}{\tensor{\overset{\wedge}{\hat{\pi}}}{_{\perp\indiq[#1]{2}}}}%
{A1m}{\tensor{\overset{\rightharpoonup}{\hat{\pi}}}{_{\indiq[#1]{1}}}}%
{A2p}{\tensor{\overset{\sim}{\hat{\pi}}}{_{\perp\indiq[#1]{2}}}}%
{A2m}{\tensor[^{\text{T}}]{\hat{\pi}}{_{\indiq[#1]{3}}}}%
}[\packageError{cosmicclass}{Unidentified Critical Case: #1}{}]%
}

\newrobustcmd{\PiPu}[2][placeholder]{%
\IfEqCase{#2}{%
{B0p}{\hat{\pi}}%
{B1p}{\tensor{\smash{\overset{\wedge}{\hat{\pi}}}}{^{\indiq[#1]{2}}}}%
{B1m}{\tensor{\smash{\hat{\pi}}}{^{\perp\indiq[#1]{1}}}}%
{B2p}{\tensor{\smash{\overset{\sim}{\hat{\pi}}}}{^{\indiq[#1]{2}}}}%
{A0p}{\tensor{\smash{\hat{\pi}}}{^\perp}}%
{A0m}{\tensor[^{\text{P}}]{\smash{\hat{\pi}}}{}}%
{A1p}{\tensor{\smash{\overset{\wedge}{\hat{\pi}}}}{^{\perp\indiq[#1]{2}}}}%
{A1m}{\tensor{\smash{\overset{\rightharpoonup}{\hat{\pi}}}}{^{\indiq[#1]{1}}}}%
{A2p}{\tensor{\smash{\overset{\sim}{\hat{\pi}}}}{^{\perp\indiq[#1]{2}}}}%
{A2m}{\tensor[^{\text{T}}]{\smash{\hat{\pi}}}{^{\indiq[#1]{3}}}}%
}[\packageError{cosmicclass}{Unidentified Critical Case: #1}{}]%
}

\newrobustcmd{\sicl}[2][placeholder]{%
\IfEqCase{#2}{%
{B0p}{\tensor{\chi}{^{\flat}}}%
{B1p}{\tensor{\smash{\overset{\wedge}{\chi}}}{^{\flat}_{\indiq[#1]{2}}}}%
{B1m}{\tensor{\chi}{^{\flat}_{\perp}_{\indiq[#1]{1}}}}%
{B2p}{\tensor{\smash{\overset{\sim}{\chi}}}{^{\flat}_{\indiq[#1]{2}}}}%
{A0p}{\tensor{\chi}{^{\flat}_\perp}}%
{A0m}{\tensor[^{\text{P}}]{\chi}{^{\flat}}}%
{A1p}{\tensor{\smash{\overset{\wedge}{\chi}}}{^{\flat}_{\perp\indiq[#1]{2}}}}%
{A1m}{\tensor{\smash{\overset{\rightharpoonup}{\chi}}}{^{\flat}_{\indiq[#1]{1}}}}%
{A2p}{\tensor{\smash{\overset{\sim}{\chi}}}{^{\flat}_{\perp\indiq[#1]{2}}}}%
{A2m}{\tensor[^{\text{T}}]{\chi}{^{\flat}_{\indiq[#1]{3}}}}%
}[\packageError{cosmicclass}{Unidentified Critical Case: #1}{}]%
}

\newrobustcmd{\ticl}[2][placeholder]{%
\IfEqCase{#2}{%
{B0p}{\tensor{\zeta}{^{\flat}}}%
{B1p}{\tensor{\smash{\overset{\wedge}{\zeta}}}{^{\flat}_{\indiq[#1]{2}}}}%
{B1m}{\tensor{\zeta}{^{\flat}_{\perp}_{\indiq[#1]{1}}}}%
{B2p}{\tensor{\smash{\overset{\sim}{\zeta}}}{^{\flat}_{\indiq[#1]{2}}}}%
{A0p}{\tensor{\zeta}{^{\flat}_\perp}}%
{A0m}{\tensor[^{\text{P}}]{\zeta}{^{\flat}}}%
{A1p}{\tensor{\smash{\overset{\wedge}{\zeta}}}{^{\flat}_{\perp\indiq[#1]{2}}}}%
{A1m}{\tensor{\smash{\overset{\rightharpoonup}{\zeta}}}{^{\flat}_{\indiq[#1]{1}}}}%
{A2p}{\tensor{\smash{\overset{\sim}{\zeta}}}{^{\flat}_{\perp\indiq[#1]{2}}}}%
{A2m}{\tensor[^{\text{T}}]{\zeta}{^{\flat}_{\indiq[#1]{3}}}}%
}[\packageError{cosmicclass}{Unidentified Critical Case: #1}{}]%
}

\newrobustcmd{\PiPl}[2][placeholder]{%
\IfEqCase{#2}{%
{B0p}{\tensor{\hat{\pi}}{^{\flat}}}%
{B1p}{\tensor{\smash{\overset{\wedge}{\hat{\pi}}}}{^{\flat}_{\indiq[#1]{2}}}}%
{B1m}{\tensor{\hat{\pi}}{^{\flat}_{\perp}_{\indiq[#1]{1}}}}%
{B2p}{\tensor{\smash{\overset{\sim}{\hat{\pi}}}}{^{\flat}_{\indiq[#1]{2}}}}%
{A0p}{\tensor{\hat{\pi}}{_\perp}^{\flat}}%
{A0m}{\tensor[^{\text{P}}]{\hat{\pi}}{^{\flat}}}%
{A1p}{\tensor{\smash{\overset{\wedge}{\hat{\pi}}}}{^{\flat}_{\perp\indiq[#1]{2}}}}%
{A1m}{\tensor{\smash{\overset{\rightharpoonup}{\hat{\pi}}}}{^{\flat}_{\indiq[#1]{1}}}}%
{A2p}{\tensor{\smash{\overset{\sim}{\hat{\pi}}}}{^{\flat}_{\perp\indiq[#1]{2}}}}%
{A2m}{\tensor[^{\text{T}}]{\hat{\pi}}{^{\flat}_{\indiq[#1]{3}}}}%
}[\packageError{cosmicclass}{Unidentified Critical Case: #1}{}]%
}

\newrobustcmd{\sic}[2][placeholder]{%
\IfEqCase{#2}{%
{B0p}{\chi}%
{B1p}{\tensor{\overset{\wedge}{\chi}}{_{\indiq[#1]{2}}}}%
{B1m}{\tensor{\chi}{_{\perp\indiq[#1]{1}}}}%
{B2p}{\tensor{\overset{\sim}{\chi}}{_{\indiq[#1]{2}}}}%
{A0p}{\tensor{\chi}{_\perp}}%
{A0m}{\tensor[^{\text{P}}]{\chi}{}}%
{A1p}{\tensor{\overset{\wedge}{\chi}}{_{\perp\indiq[#1]{2}}}}%
{A1m}{\tensor{\overset{\rightharpoonup}{\chi}}{_{\indiq[#1]{1}}}}%
{A2p}{\tensor{\overset{\sim}{\chi}}{_{\perp\indiq[#1]{2}}}}%
{A2m}{\tensor[^{\text{T}}]{\chi}{_{\indiq[#1]{3}}}}%
}[\packageError{cosmicclass}{Unidentified Critical Case: #1}{}]%
}

\newrobustcmd{\lorsicpar}[2][placeholder]{%
\IfEqCase{#2}{%
{B0m}{\tensor*[^{\text{P}}]{\smash{\underline{\chi}}}{^{\parallel}}}%
{B1p}{\tensor*{\smash{\overset{\wedge}{\chi}}}{^{\parallel}_{\indiq[#1]{2}}}}%
{B1m}{\tensor*{\chi}{^{\parallel}_{\perp\indiq[#1]{1}}}}%
{B2m}{\tensor*[^{\text{T}}]{\smash{\underline{\chi}}}{^{\parallel}_{\indiq[#1]{3}}}}%
{A0p}{\tensor*{\chi}{^{\parallel}_\perp}}%
{A0m}{\tensor*[^{\text{P}}]{\chi}{^{\parallel}}}%
{A1p}{\tensor*{\smash{\overset{\wedge}{\chi}}}{^{\parallel}_{\perp\indiq[#1]{2}}}}%
{A1m}{\tensor*{\smash{\overset{\rightharpoonup}{\chi}}}{^{\parallel}_{\indiq[#1]{1}}}}%
{A2p}{\tensor*{\smash{\overset{\sim}{\chi}}}{^{\parallel}_{\perp\indiq[#1]{2}}}}%
{A2m}{\tensor*[^{\text{T}}]{\chi}{^{\parallel}_{\indiq[#1]{3}}}}%
}[\packageError{cosmicclass}{Unidentified Critical Case: #1}{}]%
}

\newrobustcmd{\lorsicpir}[2][placeholder]{%
\IfEqCase{#2}{%
{B0p}{\tensor*{\chi}{^{\vDash}}}%
{B1p}{\tensor*{\smash{\overset{\wedge}{\chi}}}{^{\vDash}_{\indiq[#1]{2}}}}%
{B1m}{\tensor*{\chi}{^{\vDash}_{\perp\indiq[#1]{1}}}}%
{B2p}{\tensor*{\smash{\overset{\sim}{\chi}}}{^{\vDash}_{\indiq[#1]{2}}}}%
{A0p}{\tensor*{\chi}{^{\vDash}_\perp}}%
{A0m}{\tensor*[^{\text{P}}]{\chi}{^{\vDash}}}%
{A1p}{\tensor*{\smash{\overset{\wedge}{\chi}}}{^{\vDash}_{\perp\indiq[#1]{2}}}}%
{A1m}{\tensor*{\smash{\overset{\rightharpoonup}{\chi}}}{^{\vDash}_{\indiq[#1]{1}}}}%
{A2p}{\tensor*{\smash{\overset{\sim}{\chi}}}{^{\vDash}_{\perp\indiq[#1]{2}}}}%
{A2m}{\tensor*[^{\text{T}}]{\chi}{^{\vDash}_{\indiq[#1]{3}}}}%
}[\packageError{cosmicclass}{Unidentified Critical Case: #1}{}]%
}

\newrobustcmd{\lorsicper}[2][placeholder]{%
\IfEqCase{#2}{%
{B0p}{\tensor*{\chi}{^{\perp}}}%
{B1p}{\tensor*{\smash{\overset{\wedge}{\chi}}}{^{\perp}_{\indiq[#1]{2}}}}%
{B1m}{\tensor*{\chi}{^{\perp}_{\perp\indiq[#1]{1}}}}%
{B2p}{\tensor*{\smash{\overset{\sim}{\chi}}}{^{\perp}_{\indiq[#1]{2}}}}%
{A0p}{\tensor*{\chi}{^{\perp}_\perp}}%
{A0m}{\tensor*[^{\text{P}}]{\chi}{^{\perp}}}%
{A1p}{\tensor*{\smash{\overset{\wedge}{\chi}}}{^{\perp}_{\perp\indiq[#1]{2}}}}%
{A1m}{\tensor*{\smash{\overset{\rightharpoonup}{\chi}}}{^{\perp}_{\indiq[#1]{1}}}}%
{A2p}{\tensor*{\smash{\overset{\sim}{\chi}}}{^{\perp}_{\perp\indiq[#1]{2}}}}%
{A2m}{\tensor*[^{\text{T}}]{\chi}{^{\perp}_{\indiq[#1]{3}}}}%
}[\packageError{cosmicclass}{Unidentified Critical Case: #1}{}]%
}

\newrobustcmd{\Tl}[2][placeholder]{%
\IfEqCase{#2}{%
{B0p}{\tensor{\chi}{^{\flat}}}%
{B1p}{\tensor{\smash{\overset{\wedge}{\chi}}}{^{\flat}_{\indiq[#1]{2}}}}%
{B1m}{\tensor{\chi}{^{\flat}_{\perp}_{\indiq[#1]{1}}}}%
{B2p}{\tensor{\smash{\overset{\sim}{\chi}}}{^{\flat}_{\indiq[#1]{2}}}}%
{A0p}{\tensor{\chi}{^{\flat}_\perp}}%
{A0m}{\tensor[^{\text{P}}]{\mathcal{T}}{^{\flat}}}%
{A1p}{\tensor{\smash{\overset{\wedge}{\chi}}}{^{\flat}_{\perp\indiq[#1]{2}}}}%
{A1m}{\tensor{\smash{\overset{\rightharpoonup}{\mathcal{T}}}}{^{\flat}_{\indiq[#1]{1}}}}%
{A2p}{\tensor{\smash{\overset{\sim}{\chi}}}{^{\flat}_{\perp\indiq[#1]{2}}}}%
{A2m}{\tensor[^{\text{T}}]{\mathcal{T}}{^{\flat}_{\indiq[#1]{3}}}}%
}[\tensor{\mathcal{T}}{^{\flat}_{\indiq[#1]{3}}}]%
}

\newrobustcmd{\cT}[2][placeholder]{%
\IfEqCase{#2}{%
{B1p}{\tensor{\mathcal{T}}{_{\perp\indiq[#1]{2}}}}%
{B1m}{\tensor{\overset{\rightharpoonup}{\mathcal{T}}}{_{\indiq[#1]{1}}}}%
{A0m}{\tensor[^{\text{P}}]{\mathcal{T}}{}}%
{A2m}{\tensor[^{\text{T}}]{\mathcal{T}}{_{\indiq[#1]{3}}}}%
}[\packageError{cosmicclass}{Unidentified Critical Case: #1}{}]%
}

\newrobustcmd{\cTLambda}[2][placeholder]{%
\IfEqCase{#2}{%
{B1p}{\tensor{\zeta}{_{\perp\indiq[#1]{2}}}}%
{B1m}{\tensor{\overset{\rightharpoonup}{\zeta}}{_{\indiq[#1]{1}}}}%
{A0m}{\tensor[^{\text{P}}]{\zeta}{}}%
{A2m}{\tensor[^{\text{T}}]{\zeta}{_{\indiq[#1]{3}}}}%
}[\packageError{cosmicclass}{Unidentified Critical Case: #1}{}]%
}

\newrobustcmd{\cTmul}[2][placeholder]{%
\IfEqCase{#2}{%
{B1p}{\tensor{\upsilon}{_{\perp\indiq[#1]{2}}}}%
{B1m}{\tensor{\overset{\rightharpoonup}{\upsilon}}{_{\indiq[#1]{1}}}}%
{A0m}{\tensor[^{\text{P}}]{\upsilon}{}}%
{A2m}{\tensor[^{\text{T}}]{\upsilon}{_{\indiq[#1]{3}}}}%
}[\packageError{cosmicclass}{Unidentified Critical Case: #1}{}]%
}

\newrobustcmd{\cTpic}[2][placeholder]{%
\IfEqCase{#2}{%
{B1p}{\tensor{\phi}{_{\perp\indiq[#1]{2}}}}%
{B1m}{\tensor{\overset{\rightharpoonup}{\phi}}{_{\indiq[#1]{1}}}}%
{A0m}{\tensor[^{\text{P}}]{\phi}{}}%
{A2m}{\tensor[^{\text{T}}]{\phi}{_{\indiq[#1]{3}}}}%
}[\packageError{cosmicclass}{Unidentified Critical Case: #1}{}]%
}

\newrobustcmd{\cTpicl}[2][placeholder]{%
\IfEqCase{#2}{%
{B1p}{\tensor{\phi}{^{\flat}_{\perp\indiq[#1]{2}}}}%
{B1m}{\tensor{\overset{\rightharpoonup}{\phi}}{^{\flat}_{\indiq[#1]{1}}}}%
{A0m}{\tensor[^{\text{P}}]{\phi}{^{\flat}}}%
{A2m}{\tensor[^{\text{T}}]{\phi}{^{\flat}_{\indiq[#1]{3}}}}%
}[\packageError{cosmicclass}{Unidentified Critical Case: #1}{}]%
}

\newrobustcmd{\cTPiP}[2][placeholder]{%
\IfEqCase{#2}{%
{B1p}{\tensor{\varpi}{_{\perp\indiq[#1]{2}}}}%
{B1m}{\tensor{\overset{\rightharpoonup}{\varpi}}{_{\indiq[#1]{1}}}}%
{A0m}{\tensor[^{\text{P}}]{\varpi}{}}%
{A2m}{\tensor[^{\text{T}}]{\varpi}{_{\indiq[#1]{3}}}}%
}[\packageError{cosmicclass}{Unidentified Critical Case: #1}{}]%
}

\newrobustcmd{\cTl}[2][placeholder]{%
\IfEqCase{#2}{%
{B1p}{\tensor{\mathcal{T}}{^{\flat}_{\perp\indiq[#1]{2}}}}%
{B1m}{\tensor{\smash{\overset{\rightharpoonup}{\mathcal{T}}}}{^{\flat}_{\indiq[#1]{1}}}}%
{A0m}{\tensor[^{\text{P}}]{\mathcal{T}}{^{\flat}}}%
{A2m}{\tensor[^{\text{T}}]{\mathcal{T}}{^{\flat}_{\indiq[#1]{3}}}}%
}[\packageError{cosmicclass}{Unidentified Critical Case: #1}{}]%
}

\newrobustcmd{\cTu}[2][placeholder]{%
\IfEqCase{#2}{%
{B1p}{\tensor{\mathcal{T}}{^{\perp\indiq[#1]{2}}}}%
{B1m}{\tensor{\smash{\overset{\rightharpoonup}{\mathcal{T}}}}{^{\indiq[#1]{1}}}}%
{A0m}{\tensor[^{\text{P}}]{\mathcal{T}}{}}%
{A2m}{\tensor[^{\text{T}}]{\mathcal{T}}{^{\indiq[#1]{3}}}}%
}[\packageError{cosmicclass}{Unidentified Critical Case: #1}{}]%
}

\newrobustcmd{\ncTLambda}[2][placeholder]{%
\IfEqCase{#2}{%
{B0p}{\tensor{\zeta}{^{\indiq[#1]{1}}_{\indiq[#1]{1}\perp}}}%
{B1p}{\tensor{\zeta}{_{[\indiq[#1]{2}]\perp}}}%
{B1m}{\tensor{\zeta}{_{\perp\indiq[#1]{1}\perp}}}%
{B2p}{\tensor{\zeta}{_{\langle\indiq[#1]{2}\rangle\perp}}}%
}[\packageError{cosmicclass}{Unidentified Critical Case: #1}{}]%
}

\newrobustcmd{\ncTmul}[2][placeholder]{%
\IfEqCase{#2}{%
{B0p}{\tensor{\upsilon}{^{\indiq[#1]{1}}_{\indiq[#1]{1}\perp}}}%
{B1p}{\tensor{\upsilon}{_{[\indiq[#1]{2}]\perp}}}%
{B1m}{\tensor{\upsilon}{_{\perp\indiq[#1]{1}\perp}}}%
{B2p}{\tensor{\upsilon}{_{\langle\indiq[#1]{2}\rangle\perp}}}%
}[\packageError{cosmicclass}{Unidentified Critical Case: #1}{}]%
}

\newrobustcmd{\ncTpic}[2][placeholder]{%
\IfEqCase{#2}{%
{B0p}{\tensor{\phi}{^{\indiq[#1]{1}}_{\indiq[#1]{1}\perp}}}%
{B1p}{\tensor{\phi}{_{[\indiq[#1]{2}]\perp}}}%
{B1m}{\tensor{\phi}{_{\perp\indiq[#1]{1}\perp}}}%
{B2p}{\tensor{\phi}{_{\langle\indiq[#1]{2}\rangle\perp}}}%
}[\packageError{cosmicclass}{Unidentified Critical Case: #1}{}]%
}

\newrobustcmd{\ncTpicl}[2][placeholder]{%
\IfEqCase{#2}{%
  {B0p}{\tensor{\phi}{^{\flat}^{\indiq[#1]{1}}_{\indiq[#1]{1}\perp}}}%
{B1p}{\tensor{\phi}{^{\flat}_{[\indiq[#1]{2}]\perp}}}%
{B1m}{\tensor{\phi}{^{\flat}_{\perp\indiq[#1]{1}\perp}}}%
{B2p}{\tensor{\phi}{^{\flat}_{\langle\indiq[#1]{2}\rangle\perp}}}%
}[\packageError{cosmicclass}{Unidentified Critical Case: #1}{}]%
}

\newrobustcmd{\ncTPiP}[2][placeholder]{%
\IfEqCase{#2}{%
{B0p}{\tensor{\varpi}{^{\indiq[#1]{1}}_{\indiq[#1]{1}\perp}}}%
{B1p}{\tensor{\varpi}{_{[\indiq[#1]{2}]\perp}}}%
{B1m}{\tensor{\varpi}{_{\perp\indiq[#1]{1}\perp}}}%
{B2p}{\tensor{\varpi}{_{\langle\indiq[#1]{2}\rangle\perp}}}%
}[\packageError{cosmicclass}{Unidentified Critical Case: #1}{}]%
}

\newrobustcmd{\ncT}[2][placeholder]{%
\IfEqCase{#2}{%
{B0p}{\tensor{\mathcal{T}}{^{\indiq[#1]{1}}_{\indiq[#1]{1}\perp}}}%
{B1p}{\tensor{\mathcal{T}}{_{[\indiq[#1]{2}]\perp}}}%
{B1m}{\tensor{\mathcal{T}}{_{\perp\indiq[#1]{1}\perp}}}%
{B2p}{\tensor{\mathcal{T}}{_{\langle\indiq[#1]{2}\rangle\perp}}}%
}[\packageError{cosmicclass}{Unidentified Critical Case: #1}{}]%
}

\newrobustcmd{\cR}[2][placeholder]{%
\IfEqCase{#2}{%
{A0p}{\tensor{\underline{\mathcal{R}}}{}}%
{A0m}{\tensor[^{\text{P}}]{\mathcal{R}}{_{\perp\circ}}}%
{A1p}{\tensor{\underline{\mathcal{R}}}{_{[\indiq[#1]{2}]}}}%
{A1m}{\tensor{\mathcal{R}}{_{\perp\indiq[#1]{1}}}}%
{A2p}{\tensor{\underline{\mathcal{R}}}{_{\langle\indiq[#1]{2}\rangle}}}%
{A2m}{\tensor[^{\text{T}}]{\mathcal{R}}{_{\perp\indiq[#1]{3}}}}%
}[\packageError{cosmicclass}{Unidentified Critical Case: #1}{}]%
}

\newrobustcmd{\cRLambda}[2][placeholder]{%
\IfEqCase{#2}{%
{A0p}{\tensor{\underline{\zeta}}{}}%
{A0m}{\tensor[^{\text{P}}]{\zeta}{_{\perp\circ}}}%
{A1p}{\tensor{\underline{\zeta}}{_{[\indiq[#1]{2}]}}}%
{A1m}{\tensor{\zeta}{_{\perp\indiq[#1]{1}}}}%
{A2p}{\tensor{\underline{\zeta}}{_{\langle\indiq[#1]{2}\rangle}}}%
{A2m}{\tensor[^{\text{T}}]{\zeta}{_{\perp\indiq[#1]{3}}}}%
}[\packageError{cosmicclass}{Unidentified Critical Case: #1}{}]%
}

\newrobustcmd{\cRl}[2][placeholder]{%
\IfEqCase{#2}{%
  {A0p}{\tensor{\underline{\mathcal{R}}}{^{\flat}}}%
{A0m}{\tensor[^{\text{P}}]{\mathcal{R}}{^{\flat}_{\perp\circ}}}%
{A1p}{\tensor{\underline{\mathcal{R}}}{^{\flat}_{[\indiq[#1]{2}]}}}%
{A1m}{\tensor{\mathcal{R}}{^{\flat}_{\perp\indiq[#1]{1}}}}%
{A2p}{\tensor{\underline{\mathcal{R}}}{^{\flat}_{\langle\indiq[#1]{2}\rangle}}}%
{A2m}{\tensor[^{\text{T}}]{\mathcal{R}}{^{\flat}_{\perp\indiq[#1]{3}}}}%
}[\packageError{cosmicclass}{Unidentified Critical Case: #1}{}]%
}

\newrobustcmd{\cRu}[2][placeholder]{%
\IfEqCase{#2}{%
{A0p}{\tensor{\underline{\mathcal{R}}}{}}%
{A0m}{\tensor[^{\text{P}}]{\mathcal{R}}{_{\perp\circ}}}%
{A1p}{\tensor{\underline{\mathcal{R}}}{^{[\indiq[#1]{2}]}}}%
{A1m}{\tensor{\mathcal{R}}{^{\perp\indiq[#1]{1}}}}%
{A2p}{\tensor{\underline{\mathcal{R}}}{^{\langle\indiq[#1]{2}\rangle}}}%
{A2m}{\tensor[^{\text{T}}]{\mathcal{R}}{^{\perp\indiq[#1]{3}}}}%
}[\packageError{cosmicclass}{Unidentified Critical Case: #1}{}]%
}

\newrobustcmd{\ncR}[2][placeholder]{%
\IfEqCase{#2}{%
  {A0p}{\tensor{\mathcal{R}}{_{\perp\perp}}}%
{A0m}{\tensor[^{\text{P}}]{\mathcal{R}}{_{\circ\perp}}}%
{A1p}{\tensor{\mathcal{R}}{_{\perp[\indiq[#1]{2}]\perp}}}%
{A1m}{\tensor{\mathcal{R}}{_{\indiq[#1]{1}\perp}}}%
{A2p}{\tensor{\mathcal{R}}{_{\perp\langle\indiq[#1]{2}\rangle\perp}}}%
{A2m}{\tensor[^{\text{T}}]{\mathcal{R}}{_{\indiq[#1]{3}\perp}}}%
}[\packageError{cosmicclass}{Unidentified Critical Case: #1}{}]%
}

\newrobustcmd{\ncRLambda}[2][placeholder]{%
\IfEqCase{#2}{%
  {A0p}{\tensor{\zeta}{_{\perp\perp}}}%
{A0m}{\tensor[^{\text{P}}]{\zeta}{_{\circ\perp}}}%
{A1p}{\tensor{\zeta}{_{\perp[\indiq[#1]{2}]\perp}}}%
{A1m}{\tensor{\zeta}{_{\indiq[#1]{1}\perp}}}%
{A2p}{\tensor{\zeta}{_{\perp\langle\indiq[#1]{2}\rangle\perp}}}%
{A2m}{\tensor[^{\text{T}}]{\zeta}{_{\indiq[#1]{3}\perp}}}%
}[\packageError{cosmicclass}{Unidentified Critical Case: #1}{}]%
}

\newrobustcmd{\Proj}[2][placeholder]{%
\IfEqCase{#2}{%
  {A2m}{\tensor[^{\text{T}}]{\check{\mathcal{P}}}{#1}}%
}[\packageError{cosmicclass}{Unidentified Critical Case: #1}{}]%
}

\newrobustcmd{\Projl}[2][placeholder]{%
\IfEqCase{#2}{%
  {A2m}{\tensor[^{\text{T}}]{\check{\mathcal{P}}}{^{\flat}#1}}%
}[\packageError{cosmicclass}{Unidentified Critical Case: #1}{}]%
}

\newrobustcmd{\fA}{%
  {\tensor{\mathcal{  A}}{_{\acu{u}}}}%
}
\newrobustcmd{\fB}{%
  {\tensor{\mathcal{  B}}{_{\acu{v}}}}%
}
\newrobustcmd{\fC}{%
  {\tensor{\mathcal{  C}}{^{\acu{v}}}}%
}
\newrobustcmd{\fphi}{%
  {\tensor{\phi}{^{\acu{w}}}}%
}
\newrobustcmd{\fpi}{%
  {\tensor{\pi}{_{\acu{w}}}}%
}

\newrobustcmd{\covard}[2]{%
  {\frac{\bar{\delta}#1}{\bar{\delta}#2}}
}
\newrobustcmd{\copard}[2]{%
  {\frac{\bar{\partial}#1}{\bar{\partial}#2}}
}
\newrobustcmd{\pard}[2]{%
  {\frac{\partial #1}{\partial #2}}
}

\def\nar{\rm{New A Rev.}}

\def\prd{\rm{Phys.~Rev.~D}}

\def\ptrsla{\rm{Philos.~Trans.~R.~Soc~A}}

\def\jmp{\rm{J.~Math.~Phys.}}
\def\cqg{\rm{Class.~Quantum~Gravity}}

\def\rpp{\rm{Rep.~Prog.~Phys.}}
\def\ijmpd{\rm{Int.~J.~Mod.~Phys.~D}}

\def\plettb{\rm{Phys.~Lett.~B}}

\def\rmphyo{\rm{Rev.~Mod.~Phys}}

\def\progtp{\rm{Prog.~Theor.~Phys.}}
\def\appb{\rm{Acta~Phys.~Pol.~B}}

\def\ijtp{\rm{Int.~J.~Theor.~Phys.}}

\def\nar{\rm{New.~Astron.~Rev.}}

\def\ijgmmp{\rm{Int.~J.~Geom.~Meth.~Mod.~Phys.}}

\newrobustcmd{\PPM}[1]{%
  {\left[\tensor*{\mathsf{M}}{_{\ }^{\left(\text{#1}\right)}}\right]}%
}

\tikzset{
  good/.style={circle, opacity=0.7, draw=green!60, fill=green!5, line width=.8mm, minimum size=3.5mm},
  bad/.style={circle, opacity=0.7, draw=red!60, fill=red!5, line width=.8mm, minimum size=3.5mm},
  badx/.style={circle, opacity=0.4, draw=red!60, fill=red!5, line width=.8mm, minimum size=3.5mm},
  save/.style={circle, opacity=0.7, draw=green!60, fill=green!60, line width=1.5mm, minimum size=3.5mm},
  kill/.style={circle, opacity=0.7, draw=red!60, fill=red!60, line width=1.5mm, minimum size=3.5mm},
  bind/.style={draw=green!60, opacity=0.7, line width=1mm,},
  wrap/.style={draw=red!60, opacity=0.7, line width=1mm,},
}